\documentstyle[preprint,pre,eqsecnum,aps,amssymb,epsfig]{revtex}

\begin{document}
\tightenlines

\draft
\title{Wetting films on chemically heterogeneous substrates} 
\author{C. Bauer and S. Dietrich}
\address{Fachbereich Physik, Bergische Universit\"at Wuppertal,\\
D-42097 Wuppertal, Germany}
\date{June 11, 1999}

\maketitle

\begin{abstract}
Based on a microscopic density functional theory we investigate the
morphology of thin liquidlike wetting films adsorbed on substrates endowed
with well-defined chemical heterogeneities. As paradigmatic cases we
focus on a single chemical step and on a single stripe. In view of
applications in microfluidics the accuracy of guiding liquids by
chemical microchannels is discussed. Finally we give a general
prescription of how to investigate theoretically the wetting
properties of substrates with arbitrary chemical structures.
\end{abstract}


\section{Introduction}
\label{s:introduction}

A large variety of experimental techniques has emerged which allows
one to endow solid substrates with stable, persistent, and
well-defined lateral patterns of geometrical or chemical nature or a
combination thereof
\cite{hansenetal,tolfree,xiawhitesides,burmeisteretal,rogersetal,aizenbergetal}.
The spatial extension of these regular structures ranges from the $\mu$m
scale down to nm. A major part of the application potential of such
man-made surfaces is based on their exposure to fluids and their
ability to imprint permanent, well-defined structures on
adjacent soft matter encompassing simple anorganic and organic liquids
as well as complex fluids such as polymer solutions or colloidal
suspensions (see, e.g., Ref.~\cite{sdsicily}).

Here we are interested in planar, chemically structured
surfaces. Recent experiments have explored the wetting and adsorption
properties of fluids and polymer solutions on such heterogeneous
surfaces~\cite{patersonfermigier,wiegandetal,zeppenfeldetal,karimetal,tilo,rockfordetal,gauetal,gallardoetal}; 
some studies even have demonstrated the feasibility to control the
growth of biological systems by attaching them to structured
surfaces~\cite{chenetal,matsuzawa} and to recognize biological 
molecules, e.g., proteins, selectively by bringing them into contact
with nanostructured surfaces~\cite{delamarcheetal,shietal}. In the context of
microfluidics~\cite{knightetal,grunze} one is interested in guiding
small quantities of valuable liquids to designated sites where they
can either be analyzed or undergo chemical reactions. This
transportation problem can be solved by using microgrooves or
microchannels. Alternatively, liquids can be guided along chemical
lanes on flat substrates. Such an integrated network can be used to
build chemical chips and tiny chemical factories (see, e.g.,
Ref.~\cite{service}). This development has been greatly facilitated by
the emergence of microcontact
printing~\cite{wilburetal,kumaretal,jackmanetal,hidberetal} as an important
technique to cover solid substrates with designed chemical
patterns. With this technique one can create surface patterns which
consist of regular monolayer patches composed of different chemical
species. These patterns are anchored at a homogeneous and flat
substrate such that the resulting decorated surface remains flat on
the molecular scale. Such a patchwork of, e.g., hydrophilic and
hydrophobic areas with lateral extensions in the sub-micrometer range
can be designed with high precision and turns out to be rather robust.

In view of the desired miniaturization of such structures the question
arises to which extent thermal fluctuations limit the ability to keep
adsorbed liquids with high lateral precision on the designated
chemical patterns without spilling. Moreover, one would like to know
which chemical and structural architecture of the man-made surface is
favorable for miniaturization. This requires to decode the relation
between the structural properties of the highly inhomogeneous liquids
and the molecular interaction potentials of the fluid particles and of
the various, specificly arranged substrate particles.

We address some of these issues for the case of thermal equilibrium by
suitable tools of statistical mechanics. Different theoretical models
have already been applied to study various basic properties of
adsorption of liquids on structured substrates. Exact virial theorems
and compressibility sum rules, as they can be formulated for
fluids in contact with homogeneous walls, have been derived also for
heterogeneous substrates~\cite{henderson}. In this context phenomenological
interface displacement and square-gradient
models~\cite{gauetal,pomeauvannimenus,robbinsandelmanjoanny,boulter,lenzlipowsky,swainlipowsky,lipowskylenzswain,bauer2} as
well as lattice gas models~\cite{nechaevzhang,urbanetal,burkhardt}
highlight the general behavior of liquidlike wetting layers and their
singular properties at surface or morphological phase
transitions. More sophisticated theories such as density functional
theories~\cite{colevittoratos,lajtarsokolowski,chmieletal,roeckenetal,douglasfrinksalinger,nathnealeydepablo}
and computer
simulations~\cite{karimetal,gacetal,schoendiestler,bockschoen,curtaroloetal}  
have the potential to investigate in detail the role of microscopic
interactions and to resolve the fine structures of the fluids under
consideration on a molecular scale. Whereas the former theories are
based on severe simplifications and focus on idealized systems on a
macroscopic scale (i.e., $\mu$m and larger), the latter require high
computational efforts such that only rather small systems like pores
and slits can be studied. Our present analysis is supposed to provide
a link between these two approaches in that it keeps track of the
microscopic molecular interactions between the various constituents
but leaves out fine structures on the molecular scale. This allows us
to investigate liquidlike structures on a mesoscopic scale
commensurate with the actual experimental system sizes. As basic
ingredients for the molecular interactions we adopt Lennard-Jones,
i.e., long-ranged interaction potentials. These potentials are not
only applicable for rare gases but resemble also reliable effective
potentials for molecules like small alkanes or
alcohols~\cite{israelachvili}. 

In order to provide the information and terminology required for the subsequent
considerations in Sec.~\ref{s:theoryhomog} we recall some basic results for wetting
of flat and homogeneous substrates~\cite{dietrichschick,sdreview} and
extend them to the case of homogeneous substrates covered by a
laterally homogeneous overlayer. In Sec.~\ref{s:theoryinhomog} we
present the theoretical basis for our analysis of wetting of
chemically structured substrates from which 
we derive many analytical and numerical results in
Secs.~\ref{s:analytical} and 
\ref{s:numerical}, respectively. These results extend and complete
earlier studies which also focused on the mesoscopic structure of
liquidlike layers on heterogeneous
substrates~\cite{colevittoratos,koch}. Our analysis is carried out
within mean field theory without taking into account capillary wavelike
fluctuations and the strong fluctuations arising near critical points
of the fluids. We focus our analysis on three paradigmatic chemical
surface structures: a chemical step generated by two adjacent quarter
spaces filled with different materials, a chemical step generated by
two different adjacent surface layers covering a homogeneous
substrate, and a chemical stripe generated by a slab of different
material immersed in an otherwise homogeneous substrate orthogonal to
the common flat surface. Finally, in Sec.~\ref{s:general} we
extend our approach to arbitrary chemically structured substrates and
summarize our results in Sec.~\ref{s:summary}. Technical details of
the substrate potentials and of the density functional description are
presented in Appendix~\ref{s:app_potentials} and \ref{s:app_subdiv},
respectively. 

\section{Wetting of laterally homogeneous substrates}
\label{s:theoryhomog}

\subsection{Density functional theory}

Our theoretical analysis is based on the following simple~\cite{evans}
but nonetheless successful~\cite{sdreview} density functional theory
for a spatially inhomogeneous number density $\rho({\mathbf
 r})$ of fluid particles:
\begin{eqnarray}\label{e:ldafunctional}
\Omega([\rho({\mathbf r})];T,\mu) & = & \int_{\Lambda} d^3r
f_{HS}(\rho({\mathbf r}),T) + \int_{\Lambda} d^3r
[V({\mathbf r})-\mu]\rho({\mathbf r}) \nonumber\\
& & + \frac{1}{2}\int_{\Lambda} d^3r \int_{\Lambda} d^3r'
\tilde{w}(|{\mathbf r}-{\mathbf r}'|)\rho({\mathbf r})\rho({\mathbf r}'). 
\end{eqnarray}
The minimum of $\Omega[\rho({\mathbf r})]$ with respect to
$\rho({\mathbf r})$ yields the grand canonical free energy of
the fluid corresponding to a prescribed temperature $T$ and chemical
potential $\mu$. $V({\mathbf r})$ denotes the
substrate potential and $\tilde{w}({\mathbf r})$ is the attractive
part of the fluid interparticle potential.
$\Lambda$ is the volume occupied by the fluid particles.
$f_{HS}$ is the free energy density of a hard-sphere reference fluid.
This contribution takes into account the repulsive part of the fluid
interparticle potential within a local density approximation which
neglects short-ranged particle-particle correlations. These
correlations become important in the close vicinity of the substrate 
surface. Nonetheless Eq.~(\ref{e:ldafunctional}) has turned out to provide
a very useful description of wetting phenomena in the case that the
adsorbed liquidlike films 
are much thicker than the diameter $\sigma_f$ of the fluid particles.

The fluid particles are assumed to interact via a Lennard-Jones potential
\begin{equation}\label{e:lennardjones}
\phi_f(r) =
4\epsilon_f\left[\left(\frac{\sigma_f}{r}\right)^{12} -
  \left(\frac{\sigma_f}{r}\right)^6\right].
\end{equation}
We apply the Weeks-Chandler-Andersen (WCA) procedure \cite{wca} to split up this
interaction into an attractive part $\phi_{att}(r)$ and a repulsive
part $\phi_{rep}(r)$. The latter gives rise to an effective,
temperature dependent hard sphere diameter
\begin{equation}
d(T) = \int\limits_0^{2^{1/6}\sigma_f}dr\,
\left\{1-\exp\left(-\frac{\phi_{rep}(r)}{k_BT}\right)\right\} 
\end{equation}
which is inserted into the Carnahan-Starling approximation for the
free energy density of the hard-sphere fluid \cite{cs}
\begin{equation}
f_{HS}(\rho,T) =
k_BT\rho\left(\ln(\rho\lambda^3)-1+\frac{4\eta-3\eta^2}{(1-\eta)^2}\right)
\end{equation}
where $\eta=\frac{\pi}{6}\rho(d(T))^3$ is the dimensionless packing
fraction and $\lambda$ is the thermal de Broglie wavelength.

The attractive part of the interaction $\phi_{att}(r)$ is approximated
by
\begin{equation}\label{e:attractivepot}
\tilde{w}(r) = \frac{4w_0\sigma_f^3}{\pi^2}(r^2+\sigma_f^2)^{-3}
\end{equation}
with
\begin{equation}
w_0 = \int_{{\mathbb R}^3}d^3r\,\tilde{w}(r) = \int_{{\mathbb
R}^3}d^3r\,\phi_{att}(r) = -\frac{32}{9}\sqrt{2}\pi\epsilon_f\sigma_f^3.
\end{equation}
Although Eq.~(\ref{e:attractivepot}) does not strictly implement the
WCA procedure corresponding to Eq.~(\ref{e:lennardjones}), it
resembles it closely and offers the valuable advantage of reduced
computational efforts because the form of $\tilde{w}(r)$ allows one to
carry out certain integrations over $\tilde{w}(r)$ analytically.
For large particle separations $r$ one has $\tilde{w}(r\to\infty)\sim r^{-6}$;
the amplitude is chosen such that the integrated
strength equals that of the attractive contribution $\phi_{att}$ obtained by
the strict application of the WCA procedure. The third integral in
Eq.~(\ref{e:ldafunctional}) takes into account the attractive fluid
interparticle interaction within mean field theory based on
Eq.~(\ref{e:attractivepot}).

\subsection{Bulk phases}
\label{s:bulk}

In a bulk system the particle density $\rho_{\gamma}$ (where
$\gamma=l,g$ denote the liquid and vapor phase, respectively) is
spatially constant, leading to (see Eq.~(\ref{e:ldafunctional}))
\begin{equation}\label{e:bfedensity}
\Omega_b(\rho_{\gamma},T,\mu) = f_{HS}(\rho_{\gamma},T) +
\frac{1}{2}w_0\rho_{\gamma}^2 - \mu\rho_{\gamma}
\end{equation}
for the grand canonical free energy density of a bulk fluid. Minimization of
$\Omega_b$ with respect to $\rho_{\gamma}$ yields the equilibrium densities.
The conditions for liquid-vapor phase coexistence $\mu=\mu_0(T)$ are
\begin{equation}
\left.\frac{\partial\Omega_b}{\partial\rho}\right|_{\rho=\rho_g} = 
\left.\frac{\partial\Omega_b}{\partial\rho}\right|_{\rho=\rho_l} = 0
\quad \mbox{and} \quad \Omega_b(\rho_g) = \Omega_b(\rho_l).
\end{equation}
Off coexistence, i.e., for $\mu\neq\mu_0$, only the liquid or the
vapor phase is stable. In this case the density of the metastable
phase corresponds to the second local minimum of $\Omega_b$.

\subsection{Wetting of laterally homogeneous substrates}
\label{s:hom_dft}

\begin{figure}
\begin{center}
\epsfig{file=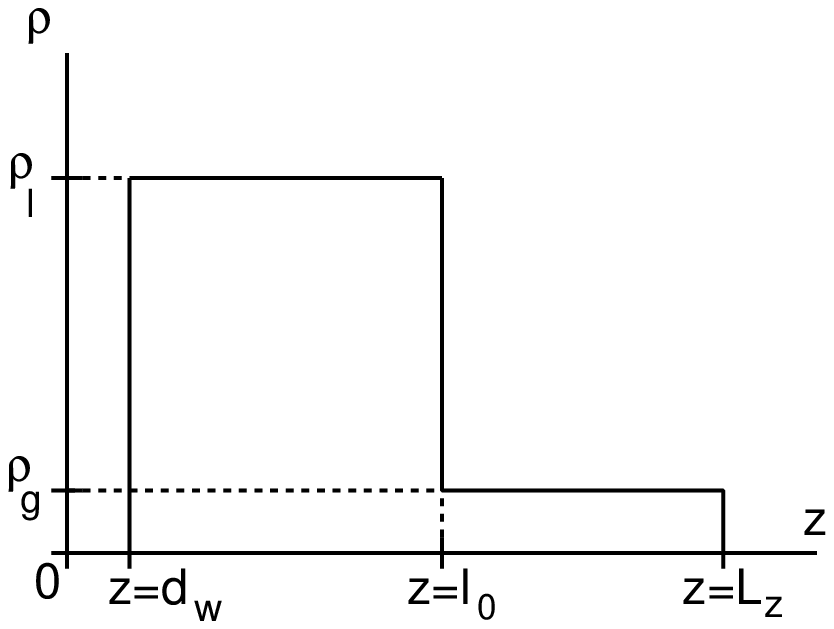, width=13cm}
\end{center}
\caption{\label{f:hom_sharpkink}
Sharp-kink approximation for the particle density distribution. At the
position $z = l_0$ of the liquid-vapor interface there is a steplike
variation of the particle density between the bulk liquid ($\rho_l$) and the
bulk vapor density ($\rho_g$). At $z=L_z$ the density is truncated
in order to facilitate the thermodynamic limit. Due to the repulsion
between the fluid and the substrate particles the particle density
vanishes for $z < d_w$ giving rise to an excluded volume.} 
\end{figure}

Here we recall some basic results for wetting phenomena (for reviews
see, e.g.,
Refs.~\cite{sdreview,degennesreview,schick,sullivantelodagama,evansnew})
as they follow from Eq.~(\ref{e:ldafunctional}) for a homogeneous
substrate within the so-called sharp-kink approximation (see
Fig.~\ref{f:hom_sharpkink} and Ref.~\cite{sdreview}). This allows us
to formulate our corresponding findings for substrates which are
heterogeneous either in the direction normal to the surface or in
lateral directions.

In the following we consider a flat substrate located in the half
space $w = \{{\mathbf r}\in{\mathbb R}^3|z\leq 0\}$. The substrate
is either homogeneous or composed of a homogeneous part $w_H =
\{{\mathbf r}\in{\mathbb R}^3|z<-dg_z\}$ covered by a homogeneous
surface layer $w_S=\{{\mathbf r}\in{\mathbb R}^3|-dg_z\leq z \leq 0\}$
of different chemical species. It consists of particles located
on an orthorhombic lattice 
with lattice constants $g_i$ in the $i$ direction ($i=x,y,z$). For
reasons of simplicity the lattice constants $g_i$ are assumed to be
constant throughout the whole substrate. The
surface layer consists of $n=d+1$ monolayers with the uppermost and
the lowest monolayer located at $z=0$ and $z=-dg_z$, respectively. The
theoretical description (Eq.~(\ref{e:ldafunctional})) is identical for
both substrate types; it only 
differs with respect to the considered substrate potential $V(z)$. 
The substrate acts as an inert spectator phase.
The interaction potential between a fluid particle and an individual
substrate particle is taken to be of the Lennard-Jones type, too:
\begin{equation}
\phi_{H,S}(r) =
4\epsilon_{H,S}\left[\left(\frac{\sigma_{H,S}}{r}\right)^{12} -
  \left(\frac{\sigma_{H,S}}{r}\right)^6\right]
\end{equation}
where $H$ and $S$ denote the molecules in $w_H$ and $w_S$,
respectively. The substrate potential $V({\mathbf r}) \equiv
V(z)=V_{att}(z)+V_{rep}(z)$ follows as a laterally averaged
pairwise sum of $\phi_{H,S}$ over all substrate particles, resulting in
\begin{eqnarray}\label{e:sl_potatt}
V_{att}(z) & = & -\frac{u_3^H}{(z+(d+1)g_z)^3} -
\frac{u_4^H}{(z+(d+1)g_z)^4} \nonumber\\
 & & - u_3^S\left(\frac{1}{z^3}-\frac{1}{(z+dg_z)^3}\right) -
u_4^S\left(\frac{1}{z^4}+\frac{1}{(z+dg_z)^4}\right) + 
{\mathcal O}(z^{-5})
\nonumber\\
 & = & -\frac{u_3^H}{z^3}-\frac{2(d+1)u_4^S-(2d+1)u_4^H}{z^4}+{\mathcal
   O}(z^{-5}), \qquad z\gg dg_z
\end{eqnarray}
for the attractive potential. In Eq.~(\ref{e:sl_potatt}) we have used
the relations $u_4^{H,S} = \frac{3}{2}g_zu_3^{H,S}$. Moreover, the coefficients
can be expressed in terms of the parameters $\epsilon_{H,S}$ and
$\sigma_{H,S}$ of the molecular interactions and the lattice spacing
$g_z$. The repulsive contribution is
\begin{eqnarray}\label{e:sl_potrep}
V_{rep}(z) & = & \frac{u_9^H}{(z+(d+1)g_z)^9} +
\frac{u_{10}^H}{(z+(d+1)g_z)^{10}} \nonumber\\
 & & + u_9^S\left(\frac{1}{z^9}-\frac{1}{(z+dg_z)^9}\right) +
 u_{10}^S\left(\frac{1}{z^{10}}+\frac{1}{(z+dg_z)^{10}}\right) +
 {\mathcal O}(z^{-11}) \nonumber\\
 & = & \frac{u_9^H}{z^9} + {\mathcal O}(z^{-10}), \qquad z\gg dg_z.
\end{eqnarray} 
In the case of a homogeneous substrate, $u_i^S=u_i^H=u_i$, $V(z)$
reduces to
\begin{equation}\label{e:hom_substpot}
V(z) = -\sum_{j\geq3}\frac{u_j}{z^j}.
\end{equation}
The limit $d\to\infty$ corresponds to
a homogeneous substrate composed of the chemical species of the
surface layer.

For the above model the full minimization of the functional
in Eq.~(\ref{e:ldafunctional}) with respect 
to the spatially inhomogeneous, smooth density $\rho(z)$ can be
carried out only numerically. However, it has turned out that the
minimization restricted to the subspace of piecewise constant density
profiles provides a surprisingly accurate \emph{analytic} account of
the corresponding effective interface
potential~\cite{softkink,softkink2}. Within this so-called sharp-kink
approximation (Fig.~\ref{f:hom_sharpkink}),
\begin{equation}\label{e:hom_sharpkink}
\hat\rho(z) = \Theta(z-d_w)[\Theta(l-z)\rho_l+\Theta(z-l)\rho_g],
\end{equation}
$\rho_l$ and $\rho_g$ are the number densities of the bulk
liquid and vapor phase, respectively (see
Subsec.~\ref{s:bulk}). For $\mu<\mu_0$ $\rho_g$ is taken to be the
actual vapor density off coexistence and $\rho_l$ is the density of
the metastable liquid phase. $\Theta$ denotes the Heaviside step 
function, and $l$ is the thickness of the adsorbed liquidlike wetting 
layer, i.e., $z=l$ is the position of the emerging liquid-vapor
interface. In terms of the actual smooth density distribution
$\rho(z)$ the position of the interface can be defined by,
e.g., $\rho(z=l) = \frac{1}{2}(\rho_l+\rho_g)$. The length $d_w$ takes
into account the excluded volume for the centers of the fluid
particles due to the repulsive part of the substrate potential.
We assume that $d_w = \frac{1}{2}(\sigma_w+\sigma_f)$. The insertion
of Eq.~(\ref{e:hom_sharpkink}) into Eq.~(\ref{e:ldafunctional}) yields
\begin{equation}\label{e:hom_subdiv}
\Omega([\hat\rho(z)];T,\mu;[\hat{w}],[V]) =
\Lambda\,\Omega_b(\rho_b;T,\mu) +
A\,\Omega_s([\hat\rho(z)];T,\mu;[\tilde{w}],[V]),
\end{equation}
i.e., a decomposition of the grand canonical functional into bulk and
surface contributions. $\Lambda$ is the volume occupied by fluid
particles. In the thermodynamic limit this is the half space
$\{{\mathbf r} = ({\mathbf r}_{\parallel},z) \in {\mathbb
R}^3|z>0\}$. $A$ is the area of the 
substrate surface. $\Omega_b$ is given by Eq.~(\ref{e:bfedensity})
whereas $\Omega_s$ yields the so-called effective interface potential
$\Omega_s(l)$: 
\begin{equation}\label{e:hom_sfe}
\Omega_s([\hat\rho(z)]) = \Omega_s(l) = \Delta\Omega_b\,l +
\sigma_{wl} + \sigma_{lg} + \omega(l).
\end{equation}
The first term in Eq.~(\ref{e:hom_sfe}) measures the cost in free energy
for forming a liquidlike wetting layer of thickness
$l$ if in the bulk the vapor is the stable phase. One has
$\Delta\Omega_b = \Delta\rho\Delta\mu + {\mathcal O}((\Delta\mu)^2)$
where $\Delta\mu = \mu_0-\mu$. With
\begin{equation}\label{e:tz}
t(z) = \int\limits_z^{\infty}dz'\int_{{\mathbb R}^2}d^2r_{\parallel}
\tilde{w}(\sqrt{r_{\parallel}^2+z'^2})
\end{equation}
as the potential of a fluid particle interacting with a
half space filled with the same fluid particles, within the above sharp-kink
approximation one has for the substrate-liquid surface tension
\begin{equation}\label{e:hom_wlstens}
\sigma_{wl} = -\frac{1}{2}\rho_l^2\int\limits_0^{\infty}dz\,t(z) +
\rho_l\int\limits_{d_w}^{\infty}dz\,V(z)-d_w\Omega_b^{(l)}
\end{equation}
and
\begin{equation}
\sigma_{lg} = -\frac{1}{2}(\Delta\rho)^2\int\limits_0^{\infty}dz\,t(z)
\end{equation}
for the liquid-vapor surface tension. The $l$-dependent part
$\omega(l)$ of the effective interface potential is given by
\begin{equation}\label{e:hom_effintpot}
\omega(l) = \Delta\rho\left\{\rho_l\int\limits_{l-d_w}^{\infty}
  dz\,t(z) - \int\limits_l^{\infty}dz V(z) \right\}.
\end{equation}
Minimization of the interfacial free energy $\Omega_s(l)$
with respect to $l$ yields the equilibrium film thickness $l_0$
and the substrate-vapor surface tension
$\sigma_{wg} = \min\limits_{\{l\}} \Omega_s(l) = \Omega_s(l_0)$.
From Young's equation one obtains for the contact angle
$\cos(\theta) = (\sigma_{wg}-\sigma_{wl})/\sigma_{lg} =
1+\omega(l_0)/\sigma_{lg}$ which is thermodynamically
well-defined only at two-phase coexistence $\Delta\mu = 0$.

From Eqs.~(\ref{e:attractivepot}) and (\ref{e:tz}) one has
$t(z) = -\sum\limits_{j\geq3}t_j\,z^{-j}$ and therefore
\begin{equation}
\omega(l) = \sum_{j\geq2}\frac{a_j}{l^j}.
\end{equation}
(Here we do not consider terms $\sim l^{-5}\ln l$ generated by van der
Waals tails in the density profiles~[60(c)] not captured
by the ansatz in Eq.~(\ref{e:hom_sharpkink}).) The coefficients of the leading
terms are $a_2 = \frac{\Delta\rho}{2}(u_3^H-t_3\rho_l)$, which is
known as the Hamaker constant, and $a_3 =
\frac{\Delta\rho}{3}(2(d+1)u_4^S-(2d+1)u_4^H-(t_4+3t_3d_w)\rho_l)$;
for a homogeneous substrate with $u_j^S=u_j^H=u_j$ one has $a_3 =
\frac{\Delta\rho}{3}(u_4-(t_4+3t_3d_w)\rho_l)$. We note that the
Hamaker constant for a homogeneous substrate covered by a homogeneous
surface layer does not depend on the properties of the surface layer
but all subdominant terms do.

The substrate is said to be \emph{completely wet} by the liquid phase
if, at coexistence $\Delta\mu = 0$, $l_0$ is infinite.
Since $\omega(l_0=\infty)=0$ one has $\sigma_{wg} =
\sigma_{wl}+\sigma_{lg}$ and $\theta=0$ for completely and
$\sigma_{wg} < \sigma_{wl}+\sigma_{lg}$ and $\theta > 0$ for partially
wet substrates. If at $T=T_w$ the thickness $l_0(T,\mu=\mu_0)$ jumps from a finite 
value for $T<T_w$ to $l_0=\infty$ for $T>T_w$ the system 
undergoes a \emph{first order wetting transition}. If the film thickness grows
continuously upon approaching the wetting temperature,
i.e., $l(T\to T_w,\mu=\mu_0)\to\infty$ 
the system exhibits a \emph{critical wetting
transition}. The necessary condition
for critical wetting is that $a_2(T)$ changes sign at $T_w$ from
$a_2(T<T_w)<0$ to $a_2(T>T_w)>0$ and $a_3(T=T_w) >0$, i.e.,
$u_4-(t_4+3t_3d_w)\rho_l(T_w) > 0$.
The critical wetting transition temperature is given implicitly by 
$u_3^H = t_3\rho_l(T_w)$, or equivalently $u_3=t_3\rho_l(T_w)$ for a
homogeneous substrate.
Even with an additional surface layer, which modifies the substrate
potential in the vicinity of the substrate surface, the transition
temperature of a critical wetting transition is
determined only by the properties of the underlying homogeneous substrate
$w_H$. This implies that two substrates which differ only with respect
to their overlayers have the same wetting transition temperature if
the wetting transition is continuous. From the above formulae one can
infer that the occurrence of a critical wetting transition on a planar
and homogeneous substrate hinges on the subdominant contribution
$\sim z^{-4}$ in the asymptotic expansion of $V(z)$ for large $z$ because
for the attractive fluid-fluid interaction as given by
Eq.~(\ref{e:attractivepot}) $t_4 = 0$ and $t_3 =
-\frac{2}{3\pi}\,w_0\sigma_f^3 > 0$. 
It is possible to fulfil the necessary condition $a_3(T_w) > 0$
for critical wetting by choosing an appropriate surface layer. 
In the case of critical wetting the film thickness diverges as $l_0(T
\nearrow T_w) = -3a_3/2a_2 \sim (T_w-T)^{-1}$. 
Irrespective of the order of the wetting transition, upon approaching
coexistence along a complete wetting isotherm one has $l_0(T>T_w,
\Delta\mu\searrow 0) = (2a_2/\Delta\Omega_b)^{1/3} \sim (\Delta\mu)^{-1/3}$.
For $T>T_w$ the Hamaker constant $a_2$ is always positive. The leading
divergence of $l_0$ for a complete wetting transition is independent
of any different surface layer covering the substrate.

\section{Models for wetting of structured substrates}
\label{s:theoryinhomog}

\subsection{Simple chemical step}
\label{s:step_theory}

As a basic element for more complicated structures we first analyze
the wetting properties of a substrate which exhibits a single,
{\it s}imple {\it c}hemical {\it s}tep (SCS), i.e., a flat substrate composed
of two adjacent quarter spaces filled with different chemical
species (see Fig.~\ref{f:step_2sl_system} with $n=\infty$). The
substrate particles occupy the half space $w = 
\{{\mathbf r}\in{\mathbb R}^3|z\leq 0\}$ and the heterogeneity 
defines the position $x = 0$. The system is translationally
invariant along the $y$ direction. The substrate
and the fluid particles interact via Lennard-Jones potentials
\begin{equation}
\phi_{\pm}(r) =
4\epsilon_{\pm}\left[\left(\frac{\sigma_{\pm}}{r}\right)^{12} -
  \left(\frac{\sigma_{\pm}}{r}\right)^6\right],
\end{equation}
where the ``$+$'' and ``$-$'' signs refer to substrate particles
located in the quarter 
spaces $w_+=\{{\mathbf r}\in{\mathbb R}^3|x>0\wedge z\leq 0\}$ and $w_-=\{{\mathbf
  r}\in{\mathbb R}^3|x<0\wedge z\leq 0\}$, respectively. The substrate
potential $V(x,z)=V_{att}(x,z)+V_{rep}(x,z)$, as obtained by a pairwise
summation over all substrate particles, is given by
Eqs.~(\ref{e:app_step_potatt}) and (\ref{e:app_step_potrep}) in
Appendix~\ref{s:app_potentials}.
In the limit $|x|\to\infty$, $V(x,z)$ asymptotically approaches the substrate
potential of the respective homogeneous substrate:
\begin{eqnarray}\label{e:step_spotasym}
V(x\to\pm\infty,z) & = & -\frac{u_3^{\pm}}{z^3} -
\frac{u_4^{\pm}}{z^4} + \frac{u_9^{\pm}}{z^9} +
\frac{u_3^+-u_3^-}{2x^3} + {\mathcal O}(x^{-4}) \nonumber\\
 & = & V_{\pm}(z) + \frac{u_3^+-u_3^-}{2x^3} + {\mathcal O}(x^{-4})
\end{eqnarray}
where the coefficients of the subdominant terms omitted here depend on $z$.
Thus the adsorbed liquidlike wetting film exhibits the asymptotic
thicknesses $l_+$ for $x\to\infty$ and $l_-$ for $x\to -\infty$ which
are determined by the properties of the homogeneous substrates ``$+$''
and ``$-$'', respectively. In general one has to take into
consideration two different wetting temperatures $T_w^+$ and $T_w^-$,
one for each of the two semi-infinite, homogeneous substrates.

With the substrate potential $V(x,z)$ given, in principle the functional
in Eq.~(\ref{e:ldafunctional}) can be minimized with respect to
the full density distribution $\rho(x,z)$. However, this can
only be done numerically and requires a huge computational effort.
Therefore we focus on low temperatures which allows us to restrict the
minimization to steplike profiles
\begin{eqnarray}\label{e:step_sharpkink}
\hat\rho(x,z) & = &
[\Theta(-x)\Theta(z-d_w^-)+\Theta(x)\Theta(z-d_w^+)] \nonumber\\
& & \times [\rho_l\Theta(l(x)-z)+\rho_g\Theta(z-l(x))].
\end{eqnarray}
The quantities $d_w^{\pm}=\frac{1}{2}(\sigma_{\pm}+\sigma_f)$ take
into account the different excluded volumes in the vicinity of the substrate
surface. Because of the rapid decay of the repulsive forces $d_w$ is
taken to vary steplike at $x=0$ between $d_w^+$ and $d_w^-$.
The approximation used in Eq.~(\ref{e:step_sharpkink}) does not capture
the fluid density oscillations very close to the substrate surface. However,
these fine structures of the particle density are expected to have only small
effects if the liquidlike wetting films are rather thick, as it is the
case in the vicinity of a critical or a complete wetting transition.
The sharp-kink approximation yields an exact prediction for the
transition temperature $T_w$ of a critical wetting transition which
remains unchanged if more sophisticated models are applied
\cite{softkink}. Although within the sharp-kink approximation the
predicted thickness of adsorbed 
liquidlike wetting layers is not quantitatively accurate this approach
is expected to capture the essential features of the wetting phenomena
as considered on the present mesoscopic scale.

The grand canonical free energy functional -- which via
Eq.~(\ref{e:step_sharpkink}) is a
functional of the function $l(x)$ that describes the local
position of the liquid-vapor interface -- can be systematically decomposed
into bulk, surface and line contributions:
\begin{eqnarray}\label{e:step_subdiv}
\Omega([\hat\rho(x,z)];T,\mu;[\tilde{w}],[V]) & = &
\Lambda\,\Omega_b(\rho_g,T,\mu)
+ A\,\Omega_s(l_{\pm};T,\mu;[\tilde{w}],[V]) \nonumber\\
& & + L_y\,\Omega_l([l(x)];T,\mu;[\tilde{w}],[V]).
\end{eqnarray}
The explicit expressions for these contributions are given in
Appendix~\ref{s:app_subdiv}. In Eq.~(\ref{e:step_subdiv}) $\Lambda =
L_xL_yL_z$ is the volume 
filled with fluid particles, $A = L_xL_y$ is the surface area of
the substrate surface, and $L_y$ is the linear extension of the
chemical step. $\Omega_b$, given by Eq.~(\ref{e:bfedensity}),
is the bulk free energy density corresponding to the stable bulk vapor
phase. The surface contribution
\begin{equation}\label{e:step_sfe}
\Omega_s(l_{\pm}) = \frac{1}{2}(\Omega_s^+(l_+)+\Omega_s^-(l_-))
\end{equation}
is the arithmetic mean of the surface free energy densities
corresponding to the substrates $w_+$ and $w_-$ covered by
liquidlike films of thickness $l_+$ and $l_-$ which 
are exposed to the potentials $V_+(z)$ and $V_-(z)$, respectively (see
Eq.~(\ref{e:hom_sfe})). $l_{\pm}$ minimizes the corresponding
Eq.~(\ref{e:hom_sfe}) with
$\rho_g$ and $\rho_l$ as the bulk vapor and the metastable
bulk liquid phase, respectively.

The line contribution $\Omega_l$ is due to the substrate
heterogeneity and reads
\begin{equation}
\Omega_l[l(x)] = \tau(d_w^{\pm},l_{\pm}) + \tilde{\omega}[l(x)].
\end{equation}
The expression $\tau(d_w^{\pm}, l_{\pm})$ does not depend on the
profile $l(x)$. Minimization of the \emph{nonlocal} functional
$\tilde{\omega}[l(x)]$ (which is given in
Eq.~(\ref{e:app_step_taudep})) yields the equilibrium liquid-vapor
interface profile $\bar{l}(x)$ and the line tension 
\begin{equation}
\tau(T,\mu) = \min_{\{l(x)\}}\Omega_l[l(x)] = \Omega_l[\bar{l}(x)].
\end{equation}
A gradient expansion of $\tilde{\omega}[l(x)]$ in leading order
leads to the \emph{local} functional $\tilde{\omega}_{loc}[l(x)]$
(see Eq.~(\ref{e:app_locfunc})) that provides a
prescription of how to express the functional expressions given by a
simple, phenomenological interface displacement model in terms of the microscopic
parameters of the underlying molecular interactions.

The Euler-Lagrange equation (ELE) following from the functional derivative
of the nonlocal functional $\tilde{\omega}[l(x)]$ is
\begin{eqnarray}\label{e:step_ele}
\left.\frac{\delta\Omega_l[l(x)]}{\delta l(x)}\right|_{l=\bar{l}} 
= \left.\frac{\delta\tilde{\omega}[l(x)]}{\delta
    l(x)}\right|_{l=\bar{l}} & = &
\Delta\Omega_b-\Delta\rho[\rho_l\,t(\bar{l}(x)-d_w^+)-V(x,\bar{l}(x))]
 + I(x,\bar{l}(x)) \nonumber\\ 
& & + \Delta\rho\rho_l\left\{\bar{t}(x,\bar{l}(x)-d_w^+) -
  \bar{t}(x,\bar{l}(x)-d_w^-)\right\} = 0
\end{eqnarray}
with $\bar{t}(x,z)$ given by Eq.~(\ref{e:app_txz}) and
\begin{equation}\label{e:ele_integral}
I(x,l(x)) \equiv (\Delta\rho)^2\int\limits_{-\infty}^{\infty}dx'
\int\limits_0^{l(x')-l(x)} dz' \, \bar{w}(x-x',z')
\end{equation}
where $\bar{w}(x,z)$ is given by Eq.~(\ref{e:app_wbar}).
The ELE is a nonlocal integral equation for the function $\bar{l}(x)$.
Within the local theory the double-integral in
Eq.~(\ref{e:ele_integral}) is replaced by a differential expression
leading to
\begin{eqnarray}\label{e:step_elelocal}
I_{loc}(l(x)) & \equiv & \frac{\sigma_{lg}l\,''(x)}{(1+(l'(x))^2)^{3/2}} = 
\Delta\Omega_b-\Delta\rho[\rho_l\,t(\bar{l}(x)-d_w^+)-V(x,\bar{l}(x))]
 \nonumber\\ 
& & + \Delta\rho\rho_l\left\{\bar{t}(x,\bar{l}(x)-d_w^+) -
  \bar{t}(x,\bar{l}(x)-d_w^-)\right\}.
\end{eqnarray}
Equation~(\ref{e:step_elelocal}) is often referred to as ``augmented
Young equation'' \cite{kagan}.

\subsection{Chemical step within a surface layer}
\label{s:2sl_theory}

The analysis in the previous subsection requires that the
substrate is composed of two adjacent halves which themselves are homogeneous along
both the $x$ and the $y$ direction. However, it is not necessary that the
substrate halves are chemically homogeneous in the $z$ direction.

This allows us to consider within the same formalism a 
homogeneous substrate $w_H$ covered by a surface layer which itself is composed
of two adjacent homogeneous layers $w_{S,\pm}=\{{\mathbf r}\in{\mathbb
  R}^3|x\gtrless 0\wedge -dg_z<z<0\}$ of different chemical species. In the
following we refer to this type of substrate as a {\it l}ayer {\it
c}hemical {\it s}tep (LCS). The two halves of the layer
meet at $x=0$ (see Fig.~\ref{f:step_2sl_system}). Both surface
layers consist of $n=d+1$ monolayers with the uppermost and the lowest monolayer
located at $z=0$ and $z=-dg_z$, respectively. This model mimics rather
closely the kind of substrate inhomogeneities generated by, e.g.,
microcontact printing (see Sec.~\ref{s:introduction}). The expression
for the substrate potential $V(x,z)$ 
of this system is rather complicated (see
Eqs.~(\ref{e:app_2sl_potatt}) and (\ref{e:app_2sl_potrep})).
As mentioned in Sec.~\ref{s:hom_dft} the leading contribution to the
attractive substrate potential is only determined by the underlying 
homogeneous substrate $w_H$. Thus in the case that both substrate
halves undergo a critical wetting transition there is only a single common
wetting temperature $T_w$ for the whole substrate because the Hamaker
constant is only determined by $w_H$. 

Far from the heterogeneity, for fixed $z$, the potential $V(x,z)$
asymptotically approaches the substrate potentials $V_{\pm}(z)$ of the
two semi-infinite, laterally homogeneous substrate halves covered by a
surface layer via
\begin{equation}\label{e:2sl_spotasym}
V(x\to\pm\infty,z) =
V_{\pm}(z)+\mbox{sign}(x)\frac{3(d+1)(u_4^+-u_4^-)}{8}\frac{1}{x^4} +
{\mathcal O}(x^{-5}).
\end{equation}
Again the coefficients of the subdominant terms omitted here depend on $z$.
From Eq.~(\ref{e:2sl_spotasym}) it follows that the wetting layer thickness $l(x)$
asymptotically approaches the constant values $l_{\pm}$ corresponding
to the respective decorated substrate.

\subsection{Chemical stripe}
\label{s:stripe_theory}

As a third system we consider a substrate exhibiting a {\it c}hemical
{\it st}ripe (CST). The stripe is realized by insertion of a slab $w_{st}=\{{\mathbf
  r}\in{\mathbb R}^3||x|<a/2\wedge z\leq 0\}$ of different chemical species
``$+$'' into an otherwise homogeneous substrate $w = \{{\mathbf r}\in{\mathbb
  R}^3|z \leq 0\}$ composed of particles denoted as ``$-$'' such that in
top view a chemical stripe is formed (see
Fig.~\ref{f:stripe_system}). The system is again translationally invariant in
the $y$ direction. The substrate potential $V(-x,z) = V(x,z)$ is
given by Eqs.~(\ref{e:app_stripe_potatt}) and (\ref{e:app_stripe_potrep}).
For $z$ fixed in the limit of large $|x|$ one has
\begin{equation}\label{e:stripe_spotasym}
V(|x|\to\infty,z) = -\frac{u_3^-}{z^3} - \frac{u_4^-}{z^4} +
\frac{u_9^-}{z^9} - \frac{a}{g_x}\,\frac{u_{4,x}^+-u_{4,x}^-}{x^4} + {\mathcal
  O}(x^{-5})
\end{equation}
(where again only the coefficients of the subdominant terms depend on
$z$) implying that the equilibrium liquid-vapor interface is also symmetric
with respect to $x=0$, i.e., $l(-x) = l(x)$, and that it asymptotically
approaches the constant value $l_-$ for $|x|\to\infty$
determined by the properties of the homogeneous, flat
substrate $w$. For a wide stripe width one has
\begin{equation}
V(x,z) = -\frac{u_3^+}{z^3} -\frac{u_4^+}{z^4} +
\frac{u_9^+}{z^9} + {\mathcal O}(a^{-3}), \qquad a \gg g_x \mbox{ and
  } |x| \ll a/2,
\end{equation}
so that for small $|x|$ the profile $l(x)$ is close to the film thickness
$l_+$ of a liquidlike film on the homogeneous and planar ``$+$''
substrate.

The sharp-kink density profile used for the analysis of the CST is
\begin{eqnarray}
\hat{\rho}(x,z) & = &
\left\{\Theta\left(|x|-\frac{a}{2}\right)\,\Theta(z-d_w^-) + 
\Theta\left(\frac{a}{2}-|x|\right)\,\Theta(z-d_w^+) \right\} \nonumber\\
& & \times \left[\Theta(l(x)-z)\rho_l + \Theta(z-l(x))\rho_g \right].
\end{eqnarray}
Inserting $\hat{\rho}(x,z)$ in Eq.~(\ref{e:ldafunctional}) yields
\begin{eqnarray}
\Omega([\hat\rho(x,z)];T,\mu;[\tilde{w}],[V]) & = &
\Lambda\,\Omega_b(\rho_g,T,\mu)
+ A\,\Omega_s(l_-;T,\mu;[\tilde{w}],[V])
\nonumber\\
& & + L_y\,\Omega_l([l(x)];T,\mu;[\tilde{w}],[V]).
\end{eqnarray}
The surface contribution $\Omega_s$ (see
Appendix~\ref{s:app_subdiv}) is determined by the
properties of the homogeneous, flat substrate $w$.
Thus in contrast to the SCS here one only has to
deal with one wetting transition on $w$ at $T=T_w^-$.
The wetting transition on the stripe part $w_{st}$ is suppressed
due to the finite lateral extension of the stripe. 

The effect of the stripe on the liquid-vapor interface is
captured by the line contribution to the free energy functional. This
contribution reads 
\begin{equation}
\Omega_l[l(x)] = \tau(d_w^{\pm},l_-) + \tilde{\omega}[l(x)].
\end{equation}
Whereas $\tau(d_w^{\pm},l_-)$ (see
Eq.~(\ref{e:app_stripe_tauindep})) is independent of $l(x)$, 
$\tilde{\omega}[l(x)]$ (see Eq.~(\ref{e:app_stripe_taudep})) is a
functional of $l(x)$. Functional differentiation of $\Omega_l[l(x)]$,
i.e., of $\tilde{\omega}[l(x)]$, yields the ELE 
\begin{eqnarray}\label{e:stripe_ele}
\Delta\Omega_b & - & \Delta\rho[\rho_l\,t(\bar{l}(x)-d_w^+)-V(x,\bar{l}(x))]
- \Delta\rho\,\rho_l \int\limits_{-a/2}^{a/2}dx'
\int\limits_{\bar{l}(x)-d_w^+}^{\bar{l}(x)-d_w^-}dz\,\bar{w}(x-x',z)
\nonumber\\ 
& & + (\Delta\rho)^2\int\limits_{-\infty}^{\infty}dx'
\int\limits_0^{\bar{l}(x')-\bar{l}(x)} dz' \, \bar{w}(x-x',z') = 0.
\end{eqnarray}
The corresponding ELE within the local approximation
$\tilde{\omega}_{loc}[l(x)]$ of $\tilde{\omega}[l(x)]$ is 
\begin{eqnarray}\label{e:stripe_eleloc}
\Delta\Omega_b & - &
\Delta\rho[\rho_l\,t(\bar{l}(x)-d_w^+)-V(x,\bar{l}(x))]
\nonumber\\
& & - \Delta\rho\,\rho_l \int\limits_{-a/2}^{a/2}dx'
\int\limits_{\bar{l}(x)-d_w^+}^{\bar{l}(x)-d_w^-}dz\,\bar{w}(x-x',z)
- \frac{\sigma_{lg}\bar{l}\,''(x)}{(1+(\bar{l}\,'(x))^2)^{3/2}} = 0.
\end{eqnarray}

In the following we always discuss the actual equilibrium liquid-vapor
interface profiles; therefore we omit the overbar.

\section{Properties of the interfacial profiles}
\label{s:analytical}

\subsection{Curvature behavior}
\label{s:curvatures}

\begin{figure}
\begin{center}
\epsfig{file=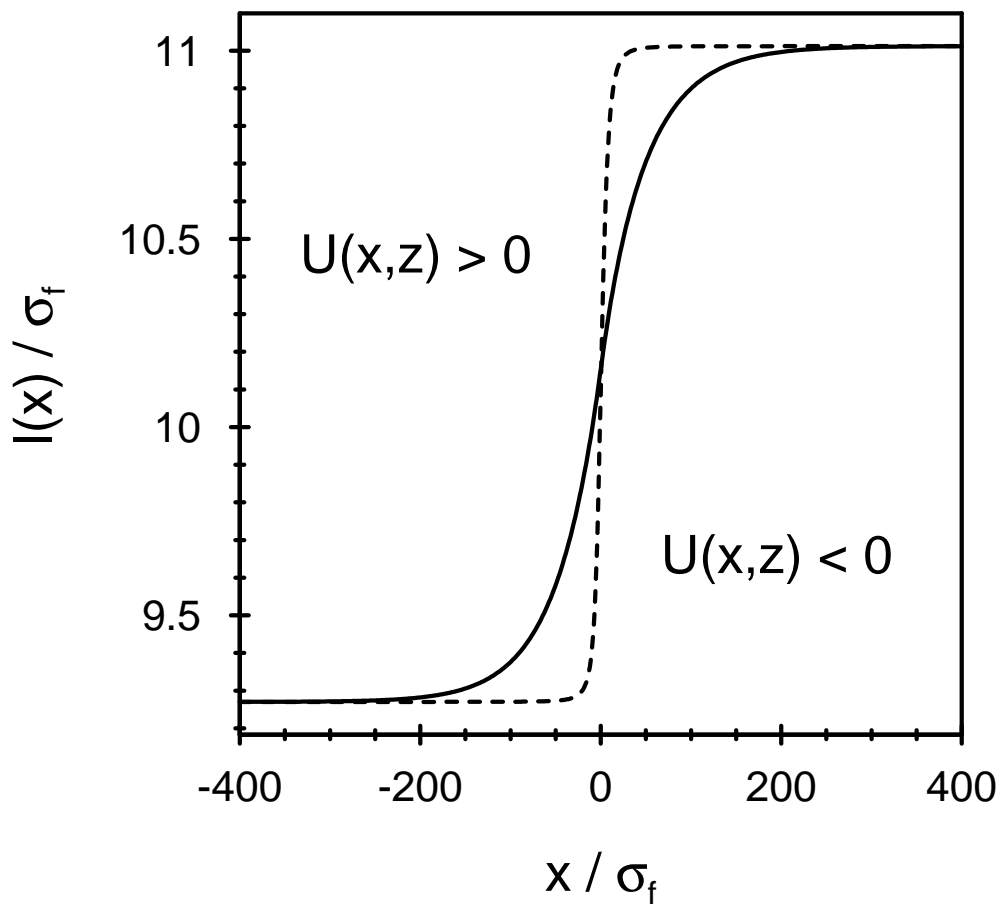, width=13cm, bbllx= 70,
  bblly=310, bburx=530, bbury=780}
\end{center}
\caption{\label{f:step_zeroline}
Typical example for the shape of a liquid-vapor interface of a
liquidlike film covering a SCS or LCS (full line) and the
corresponding contour line $U(x,z) = 0$ (dashed line) which separates regions
with positive interface curvature ($K(x,z) = U(x,z)/\sigma_{lg} > 0$)
from regions with negative curvature ($K(x,z)<0$) (see
Eqs.~(\ref{e:curvature}) and (\ref{e:curvaturepot})). The intersection
between both lines is the turning point of the interface profile. The 
interface profile
shown here is calculated for $T^*=k_BT/\epsilon_f=1.3$ and
$\Delta\mu^*=\Delta\mu/\epsilon_f=10^{-3}$ and is also displayed in
Fig.~\ref{f:step_complete}. The substrate is a SCS with the
interaction parameters chosen such that the system exhibits critical wetting
transitions at $T_w^{+*}=1.0$ and $T_w^{-*}=1.2$ on $w_+$ and $w_-$,
respectively (compare Fig.~\ref{f:step_complete}). Therefore along the
complete wetting isotherm $(T^*=1.3,\Delta\mu)$ the functions
$\Delta\Omega_b\,l + \omega_{\pm}(l)$
corresponding to the homogeneous and flat substrates $w_+$ and $w_-$
each exhibit only one extremal point and there is only a single contour line
$U(x,z)=0$ interpolating between these two extremal points.}
\end{figure}

Within the local theory the ELEs for the SCS, the LCS, and the CST
determine the local curvature $K(x)$ of the planar curve $(x,l(x))$: 
\begin{equation}\label{e:curvature}
\sigma_{lg}K(x) = \frac{\sigma_{lg}l\,''(x)}{(1+(l'(x))^2)^{3/2}} = U(x,l(x))
\end{equation}
with
\begin{equation}\label{e:curvaturepot}
U(x,z) = \Delta\Omega_b +
  \frac{\partial\omega(x,z;d_w^{\pm})}{\partial z} + {\mathcal
  C}(x,z;d_w^{\pm}).
\end{equation}
The ``local'' effective interface potential $\omega(x,z;d_w^{\pm})$ is given by
Eq.~(\ref{e:app_localeffintpot}). The function ${\mathcal
  C}(x,z;d_w^{\pm})$ takes into account the difference between $d_w^+$
and $d_w^-$, i.e., it vanishes for $d_w^+=d_w^-$ and for $d_w^+\neq
d_w^-$ in the limit $|x|\to\infty$ (see
Eqs.~(\ref{e:step_elelocal}) and (\ref{e:stripe_eleloc})). In the
limit $|x|\to\infty$ the lines $z(x)$ defined implicitly by 
$U(x,z)=0$ asymptotically approach those values for which the function
$\Delta\Omega_b\,l+\omega_{\pm}(l)$ (compare
Eq.~(\ref{e:app_step_effintpot})) is extremal with respect to $l$. One
of these values corresponds to the global minimum and thus is the
equilibrium film thickness $l_{\pm}$. In addition to this one there may be
more extremal values depending on whether the system is at or off
coexistence and on the type of wetting transition under
consideration. Since $l(x)$ asymptotically
approaches the values $l_{\pm}$ for $|x|\to\infty$ it also
approaches the contour line given by $U(x,z)=0$ corresponding to the
global minimum of 
$\Delta\Omega_b\,l+\omega_{\pm}(l)$. Figure~\ref{f:step_zeroline}
shows an example for the line $U(x,z)=0$ on a SCS and the corresponding interfacial
profile $l(x)$ for a thermodynamic state along a complete
wetting isotherm. The sign of the curvature
is given by the sign of $U(x,z)$. The point where 
$l(x)$ and the line $U(x,z)=0$ intersect is the turning point of the
profile. Due to this curvature behavior the function $l(x)$ for a SCS
or a LCS is monotonous. Analogously, for a CST $l(x)$ is monotonous
in each of the intervals $x<0$ and $x>0$.

In general the curvature behavior as predicted by the actual nonlocal
theory is different from that obtained within the local
approximation. However, in Ref.~\cite{bauer1} we have demonstrated that the 
differences between the local and the nonlocal results are negligible if the
local curvature is small, as it is the case for complete and
critical wetting transitions. Therefore, although the behavior of the
integral expression in Eqs.~(\ref{e:step_ele}) and
(\ref{e:stripe_ele}) is not analytically transparent, we expect the curvature
behavior of the profiles upon approaching a critical or complete
wetting transition to be practically the same within the nonlocal and the local
theory. 

\subsection{Asymptotic behavior of the interface profiles}
\label{s:vdwt}

The asymptotic form of the substrate potential $V(|x|\to\infty,z)$ of
a heterogeneous substrate gives rise to a behavior
\begin{equation}\label{e:vdwt}
l(|x|\to\infty) = l_{\pm}+\delta l(x) \quad \mbox{with} \quad \delta l(x) =
\lambda\,x^{-\alpha}
\end{equation}
of the interface profile
with a characteristic exponent $\alpha$. $\lambda$ is the amplitude of
the so-called van der Waals tails (VDWT) $\delta l(x)$ which
can be determined analytically for all three substrate types
SCS, LCS, and CST for those cases in which parts of the substrate,
considered per se in the corresponding homogeneous limit,
undergo a critical $(T\nearrow T_w^{\pm},\Delta\mu=0)$
or a complete $(T > T_w^{\pm},\Delta\mu\searrow 0)$ wetting transition.
These results are based on the assumption that the system is
sufficiently close to the wetting transition temperature or to
two-phase coexistence such that $l(x) \gg d_w^{\pm}$ and thus
\begin{equation}\label{e:ele_simplified}
\Delta\Omega_b + \left.\frac{\partial\omega(x,l)}{\partial
  l}\right|_{l=l(x)} = I_{(loc)}(x,l(x))
\end{equation}
for all three substrate types. $\omega(x,l)$ is given by
Eq.~(\ref{e:app_localeffintpot}) omitting the arguments $d_w^{\pm}$.

Since with $l(x) \gg d_w^{\pm}$ also
$l_{\pm} \gg d_w^{\pm}$ Eq.~(\ref{e:ele_simplified}) can be
used for inserting the ansatz $l(x) = l_{\pm} + \delta
l(x)$ into its local version. Considering then the behavior of
Eq.~(\ref{e:ele_simplified}) for large $|x|$ and exploiting the fact
that $l_{\pm}$, which corresponds to the limit $|x|\to\infty$,
minimizes Eq.~(\ref{e:app_step_sfe}) leads to the following expansion:
\begin{equation}\label{e:vdwt_eq}
a\,\delta l(x) + b\,x^{-\alpha} = \sigma_{lg}\,\delta l''(x),
\qquad \alpha > 0, 
\end{equation}
with coefficients $a$ and $b$. The term $\sim x^{-\alpha}$ stems from
the leading term in $V(|x|\to\infty)-V_{\pm}(z)$ (see
Eqs.~(\ref{e:step_spotasym}),(\ref{e:2sl_spotasym}), and 
(\ref{e:stripe_spotasym})) which determines the value of $\alpha$. The
leading asymptotic behavior of the solution of Eq.~(\ref{e:vdwt_eq})
is given by
\begin{equation}\label{e:vdwt_solution_loc}
\delta l(x) = - \frac{b}{a}\,x^{-\alpha}.
\end{equation}
Within the nonlocal theory the right-hand side is
replaced by the leading order in the series expansion of the integral
$I(x,l_{\pm}+\delta l(x))$:
\begin{equation}\label{e:vdwt_eq_nloc}
a\,\delta l(x) + b\,x^{-\alpha} = I(x,l_{\pm}+\delta l(x))
\sim x^{-\beta}, \qquad \beta > 0.
\end{equation}
Thus the asymptotic solution $\delta l(x)$ in
Eq.~(\ref{e:vdwt_eq_nloc}) for the VDWT is
equal to that obtained within the local theory
(Eq.~(\ref{e:vdwt_solution_loc})) if the right-hand side in
Eq.~(\ref{e:vdwt_eq_nloc}) 
is subdominant as compared with the left-hand side, i.e., if $\beta>\alpha$.
For $\beta = \alpha$ the amplitude $\lambda$ (see Eq.~(\ref{e:vdwt}))
obtained from the local theory differs from that following from the
nonlocal one. It turns out that this is the case
for the LCS. If one would have $\beta<\alpha$ which, however, is not
the case here, even the exponent of the power-law decay of $\delta
l(x)$ as obtained within the 
nonlocal theory would be different from that within the local theory, i.e., the
VDWT would also differ qualitatively.

With the definition $l(x\to\pm\infty) = l_{\pm}\mp\delta l_{\pm}
(x)$ for the SCS it follows that
\begin{equation}\label{e:step_vdwtcrit}
\delta l_{\pm}(x) = \lambda_{crit}^{\pm}\,x^{-3} \quad \mbox{with} \quad
\lambda_{crit}^{\pm} = \pm \frac{l_{\pm}^4}{2}\frac{a_2^+-a_2^-}{a_2^{\pm}} 
\end{equation}
in the case of a critical wetting transition on the semi-infinite, homogeneous
substrate $w_+$ or $w_-$. Along a complete wetting isotherm the result is
the same up to a numerical factor:
\begin{equation}\label{e:step_vdwtcomp}
\delta l_{\pm}(x) = \lambda_{comp}^{\pm}\,x^{-3} \quad \mbox{with} \quad
\lambda_{comp}^{\pm} = \pm \frac{l_{\pm}^4}{6}\frac{a_2^+-a_2^-}{a_2^{\pm}}.
\end{equation}
Both Eq.~(\ref{e:step_vdwtcrit}) and Eq.~(\ref{e:step_vdwtcomp}) are valid
within the local as well as the nonlocal theory in agreement with
Ref.~\cite{koch}. For the LCS one finds a different power law which
reflects the fact that for this type of substrate the lateral
dependence of the potential enters only into the subdominant terms: 
\begin{equation}\label{e:2sl_vdwt}
\delta l(x) = \lambda\,x^{-4}
\end{equation}
with
\begin{equation}\label{e:2sl_vdwtampcrit}
\lambda_{crit}^{\pm} = \left\{
\begin{array}{cc}
  \frac{9}{32}\frac{l_{\pm}^4}{|a_2|}(a_3^+-a_3^-)
  \left(1-\frac{9}{2}\,t_3\frac{(\Delta\rho)^2}{|a_2|}\right) \qquad &
  \mbox{(nonlocal)} \\
  \frac{9}{32}\frac{l_{\pm}^4}{|a_2|}(a_3^+-a_3^-) \qquad & \mbox{(local)}
\end{array} \right.
\end{equation}
and
\begin{equation}\label{e:2sl_vdwtcomp}
\lambda_{comp}^{\pm} = \left\{
\begin{array}{cc}
  \frac{3}{32}\frac{l_{\pm}^4}{a_2}(a_3^+-a_3^-)
  \left(1-3\,t_3\frac{(\Delta\rho)^2}{a_2}\right) \qquad &
  \mbox{(nonlocal)} \\
  \frac{3}{32}\frac{l_{\pm}^4}{a_2}(a_3^+-a_3^-) \qquad & \mbox{(local)}
\end{array} \right. .
\end{equation}
Both for a complete and a critical wetting transition the local and the
nonlocal theory yield the same power-law behavior but different amplitudes.
Analogously, for the CST one finds $l(|x|\to\infty) = l_-+\delta l(x)$ where
\begin{equation}\label{e:stripe_vdwtcrit}
\delta l(x) = \lambda_{crit}\,x^{-4} \qquad \mbox{with} \qquad
\lambda_{crit} = -\frac{3a}{2g_x}\,l_-^4\,\frac{a_3^+-a_3^-}{a_2^-}
\end{equation}
for critical wetting and
\begin{equation}\label{e:stripe_vdwtcomp}
\delta l(x) = \lambda_{comp}\,x^{-4} \qquad \mbox{with} \qquad
\lambda_{comp} = \frac{a}{2g_x}\,l_-^4\,\frac{a_3^+-a_3^-}{a_2^-}
\end{equation}
for a complete wetting transition on the substrate $w$. Thus for the
LCS the VDWT do resolve the difference between the local and the
nonlocal theory, but not for the SCS and the CST.

On a laterally homogeneous and flat substrate the film
thickness is given by $l_{\pm} = 3a_3^{\pm}/2|a_2^{\pm}|$ for 
critical wetting and by $l_{\pm} \sim (a_2^{\pm}/\Delta \mu)^{1/3}$
for complete wetting. For the derivation of the expressions for the
VDWT we have used these relations which do not hold for $a_3^{\pm} =
0$ in the case of critical wetting and $a_2^{\pm} = 0$ in the case of
complete wetting, respectively. In these cases of complete wetting
at $T_w$ or of tricritical wetting higher-order terms have to be taken 
into account in order to determine the amplitudes of the VDWT.
Finally we note that due to the behavior of the curvature of $l(x)$ -- as stated in
Subsec.~\ref{s:curvatures} -- the signs of the amplitudes of the VDWT
are fixed by $\mbox{sign}(l_+-l_-)$.

\subsection{Partial versus complete wetting of a chemical step}

\begin{figure}
\begin{center}
\rotatebox{-90}{\epsfig{file=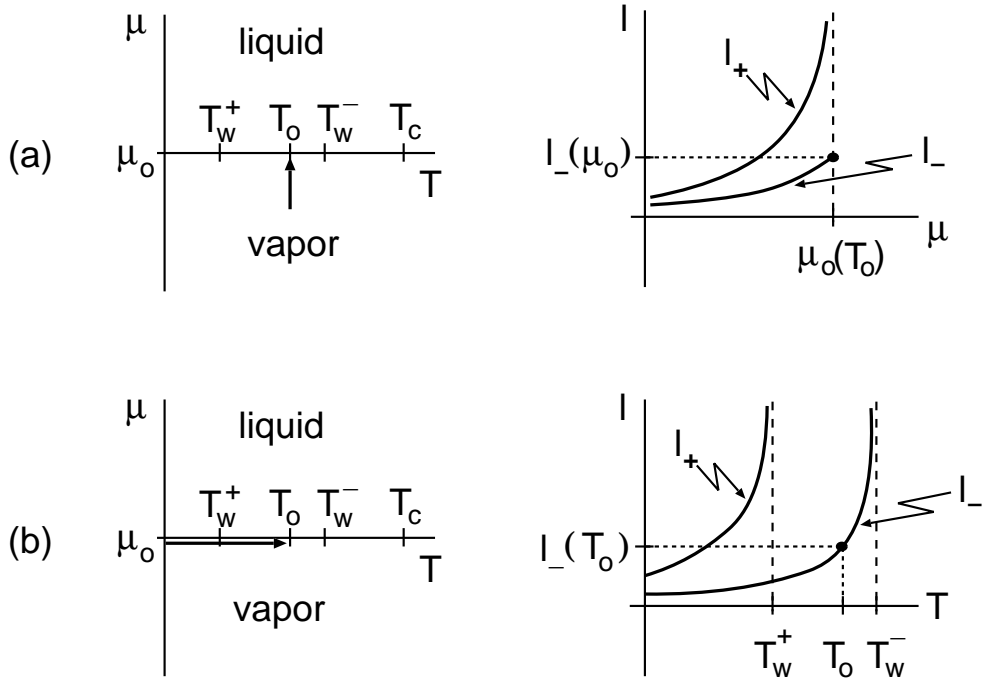, height=13cm}}
\end{center}
\caption{\label{f:step_partvscomp}
Incomplete wetting of the substrate half $x<0$ and complete wetting of the
substrate half $x>0$ can be obtained by approaching a thermodynamic state
$(T_w^+ < T_0 < T_w^-, \Delta\mu=0)$ on different thermodynamic paths
in the $(T,\mu)$ plane. This is illustrated by the corresponding
schematic behavior of the equilibrium film thickness $l_{\pm}$. For
reasons of simplicity the coexistence curve $\mu_0(T)$ is straightened
out. (a) Upon approaching $\mu_0(T_0)$ along the isotherm with $T_w^+ < T_0 <
T_w^-$ $l_+$ diverges whereas $l_-$ remains finite. (b) If the
temperature is increased along coexistence $\mu=\mu_0$ 
as $T$ approaches $T_w^+$ $l_+$ diverges and 
is macroscopicly large for $T_w^+ \leq T \leq T_0$. $l_-$ also increases
but remains finite since $T \leq T_0 < T_w^-$. Both thermodynamic
paths lead to the same final state $(T_0,\mu_0(T_0))$ of the system 
and thus to the same interfacial profile $l(x;T_0,\mu_0(T_0))$. This is
also true if the wetting transition at $T_w^+$ is first order; in this case
in (b) $l_+$ would jump to infinity at $T=T_w^+$.}
\end{figure}

The SCS and LCS allow one to realize thermodynamic states
for which one half of the substrate, say $w_+$, is completely wet whereas the
other half is only partially wet. Such a state is realized for $T_w^+ < T_0 < 
T_w^-$ and $\Delta\mu = 0$. The corresponding interface profile does
not depend on whether this state
is reached along coexistence $(T \nearrow T_0, \Delta\mu=0)$
or along a complete wetting isotherm $(T = T_0, \Delta\mu \searrow 0)$
(see Fig.~\ref{f:step_partvscomp}). The liquid-vapor interface
attains a finite value $l_-$ for $x\to -\infty$ and diverges for
$x\to\infty$.

\begin{figure}
\begin{center}
\epsfig{file=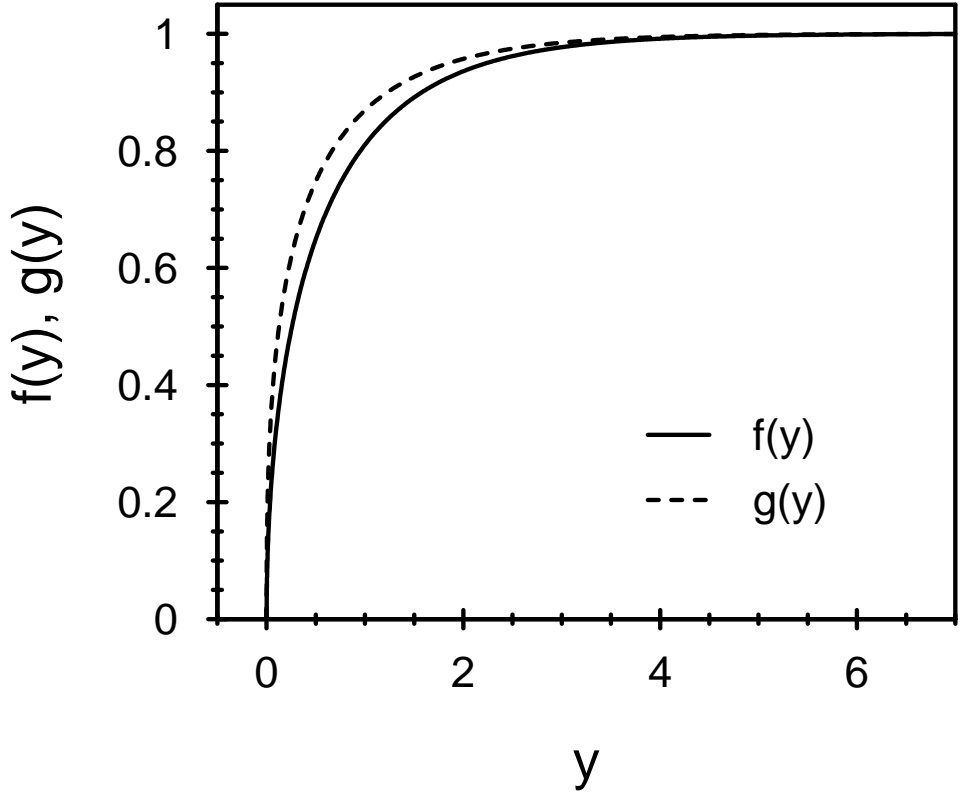, width=13cm, bbllx=65,
  bblly=330, bburx=515, bbury=715}
\end{center}
\caption{\label{f:step_scalingfcts}
Scaling functions $f$ and $g$ as defined by
Eqs.~(\ref{e:step_scalingcomp}) and (\ref{e:step_scalingcrit})
calculated for Lennard-Jones potentials ($\zeta=6$). The
scaling function $f$ governs the evolution of the interface profile if
the thermodynamic state $(T_w^+<T_0<T_w^-,\Delta\mu=0)$ at 
two-phase coexistence is approached along an isotherm
$(T=T_0,\Delta\mu)$; for this limiting thermodynamic state the 
substrate half $x>0$ of a SCS or LCS is completely wet whereas the
substrate half $x<0$ is only partially wet (see
Fig.~\ref{f:step_partvscomp}a). Analogously, the
evolution of the interfacial profile upon approaching a
critical wetting transition at $(T_0=T_w^+<T_w^-,\Delta\mu=0)$ along a
path at coexistence as shown in 
Fig.~\ref{f:step_partvscomp}b is described by the scaling function
$g$. In the limit $y\to0$ one has $f(y\to0)\sim y^{1/2}$ and
$g(y\to0)\sim y^{2/5}$. $f(y)$ and $g(y)$ attain the value $1$
for $y\to\infty$ according to $f(y\to\infty) =
1-6(\sqrt{3}-1)(\sqrt{3}+1)^{-1}e^{-3+\sqrt{3}}\,e^{-y}+{\mathcal O}(e^{-2y})$ and
$g(y\to\infty) = 1-6e^{-3}\,e^{-y}+{\mathcal O}(e^{-2y})$.}
\end{figure}

The characteristic length scale for lateral variations of the
interfacial profile $l(x)$ is the lateral height-height correlation
length $\xi_{\parallel}$. For complete wetting of the substrate
$w_{\pm}$ this length is given by (compare Ref.~\cite{lipowsky})
\begin{equation}\label{e:step_corrlencomp}
\xi_{\parallel,comp}^{\pm} = \sqrt{\sigma_{lg}}
\left(\left.\frac{d^2\omega_{\pm}}{dl^2}\right|_{l=l_{\pm}}\right)^{-1/2}
\,_{\longrightarrow \atop l_{\pm}\to\infty}\,
\sqrt{\frac{\sigma_{lg}}{6a_2^{\pm}}}\,l_{\pm}^2 \sim 
(\Delta\mu)^{-2/3}
\end{equation}
whereas for a critical wetting transition
\begin{equation}\label{e:step_corrlencrit}
\xi_{\parallel,crit}^{\pm} = \sqrt{\frac{\sigma_{lg}}{3a_3^{\pm}}}\, l_{\pm}^{5/2}
\sim t_w^{-5/2}
\end{equation}
where $t_w = (T_w-T)/T_w$ (compare Ref.~\cite{nightingaleetal}).
In the case that the substrate half $w_+$ undergoes complete and the
substrate half $w_-$ incomplete wetting we make the scaling ansatz 
\begin{equation}\label{e:step_scalingcomp}
l(x,\Delta\mu) = l_+(\Delta\mu)f(x/\xi_{\parallel,comp}^+(\Delta\mu))
\end{equation}
for the evolution of the profile 
$l(x,\Delta\mu>0)$ from a shape which attains the constant value $l_+$ in the
limit $x\to\infty$ for $\Delta\mu>0$ to the shape for $\Delta\mu=0$ with
$l(x\to\infty,\Delta\mu=0)=\infty$. With this ansatz and in the case
that the slope of the profile is small such that the
left-hand side of Eq.~(\ref{e:step_elelocal}) can be replaced by its
lowest-order expansion $\sigma_{lg}l''(x)$ one obtains the differential
equation
\begin{equation}\label{e:step_deforcomp}
1-f(y)^{-(\zeta-3)} = (\zeta-3)f''(y)
\end{equation}
for the scaling function $f(y)$ in the case that the attractive fluid-fluid
and substrate-fluid interactions decay as $r^{-\zeta}$;
$f(y\to\infty) = 1$ independent of $\zeta$. This demonstrates that the
above scaling ansatz does indeed hold. In the limit $\Delta\mu\searrow0$
which corresponds to $\xi_{\parallel,comp}^+\to\infty$, i.e., $y\to0$,
from the asymptotic behavior of the solution of
Eq.~(\ref{e:step_deforcomp}) it follows that
\begin{equation}\label{e:step_partcompcomp}
l(x\to\infty) = \gamma_{comp}\,x^{2/(\kappa+1)} \quad \mbox{with}
\quad \gamma_{comp} =
\left(\frac{(\kappa+1)^2\,a_{\sigma}^+}{2\sigma_{lg}}\right)^{1/(\kappa+1)}.
\end{equation}
The coefficients in Eq.~(\ref{e:step_partcompcomp}) are defined by the
asymptotic behavior of the effective interface potential
$\omega_+(l\to\infty) = 
a_{\sigma}^+\,l^{-\sigma}+a_{\kappa}^+\,l^{-\kappa}+{\mathcal
  O}(l^{-\kappa-1})$ where $\sigma=\zeta-4$ and $\kappa=\zeta-3$. For
Lennard-Jones potentials $\sigma=2$ and $\kappa=3$ so that
\begin{equation}\label{e:step_partcompcompvdw}
l(x\to\infty) =
\left(\frac{8\,a_2^+}{\sigma_{lg}}\right)^{1/4}\,x^{1/2}, \quad
\Delta\mu=0,\,T_w^+<T<T_w^-.
\end{equation}
This power law has been predicted originally 
by de Gennes~\cite{degennesreview}, the expression for the amplitude
is in accordance with
Ref.~\cite{koch}. Figure~\ref{f:step_scalingfcts} shows the
scaling function $f(y)$ for $\zeta=6$.

For the case $T\nearrow T_0=T_w^+$ with the critical wetting
transition temperature $T_w^+$ on the substrate half $w_+$ and
$\Delta\mu=0$ the scaling ansatz  
\begin{equation}\label{e:step_scalingcrit}
l(x\to\infty,T) = l_+(T)g(x/\xi_{\parallel,crit}^+(T))
\end{equation}
leads to the differential equation
\begin{equation}\label{e:step_deforcrit}
g(y)^{-(\zeta-3)} - g(y)^{-(\zeta-2)} = g''(y)
\end{equation}
for the scaling function $g(y)$ with $g(y\to\infty)=1$ independent of
$\zeta$. In the limit $T\nearrow T_w^+$ $\xi_{\parallel,crit}^+(T)$
diverges and from the behavior of the solution of
Eq.~(\ref{e:step_deforcrit}) for $y\to0$ one finds
\begin{equation}\label{e:step_partcompcrit}
l(x\to +\infty) = \gamma_{crit}\,x^{2/(\kappa+2)} \quad \mbox{with}
\quad \gamma_{crit} =
\left(\frac{(\kappa+2)^2\,a_{\kappa}^+}{2\sigma_{lg}}\right)^{1/(\kappa+2)}
\end{equation}
for $T=T_w^+$ so that for $\zeta=6$
\begin{equation}\label{e:step_partcompcritvdw}
l(x\to\infty) =
\left(\frac{25\,a_3^+}{2\,\sigma_{lg}}\right)^{1/5}\,x^{2/5}, \quad
\Delta\mu=0,\,T=T_w^+<T_w^-.
\end{equation}
Figure~\ref{f:step_scalingfcts} shows the
scaling function $g(y)$ for $\zeta=6$.

All these equations hold both for the SCS with $a_2^+ =
\frac{\Delta\rho}{2}(u_3^+-t_3\rho_l)$ and $a_3^+ =
\frac{\Delta\rho}{3}(u_4^+-(t_4+3t_3d_w)\rho_l)$ and for the LCS
with $a_2^+ = \frac{\Delta\rho}{2}(u_3^H-t_3\rho_l)$ and $a_3^+ =
\frac{\Delta\rho}{3}(2(d+1)u_4^+-(2d+1)u_4^H-(t_4+3t_3d_w)\rho_l)$.
The scaling functions $f$ and $g$ are independent of the
thermodynamical parameters $T$ and $\mu$ and of the amplitudes of the
molecular interactions. These dependences are completely absorbed into
$\xi_{\parallel}^+$ and $l_+$. However, $f$ and $g$ do depend on the
exponent $\zeta$ of the power-law decay of the attractive
interparticle potentials. We note that the scaling functions $f$ and
$g$ do not take into account the lateral 
long-ranged VDWT of $l(x)$ (see Subsec.~\ref{s:vdwt}). These VDWT give rise
to additional small corrections to the behavior
of $f(y\to\infty)$ and $g(y\to\infty)$.

\subsection{Partial versus complete wetting of a chemical stripe}

A similar analytic calculation as in the previous subsection can be
carried out for the case of a CST with the parameters chosen such that
at coexistence $\Delta\mu=0$ and a fixed temperature $T_w^+<T<T_w^-$
the outer area is only partially wet whereas the stripe region is
completely wet in the limit $a=\infty$. It turns out that the analytical
solution for the shape of the profile in the case of large $a$ is a
semi-ellipse: 
\begin{equation}\label{e:stripe_ellipse}
l(x;a\to\infty) =
\sqrt{\frac{2}{a}}\left(\frac{2a_2^+}{\sigma_{lg}}\right)^{1/4}
\sqrt{\frac{a^2}{4}-x^2}, 
\end{equation}
i.e., the half axes are $r_x = a/2$ and $r_z =
(2a_2^+/\sigma_{lg})^{1/4}\sqrt{a/2}$. The corresponding excess
coverage, in comparison with the case $a=0$, scales as
$\Gamma_{ex}(a\to\infty) \sim a^{3/2}$. In the 
limit $a\to\infty$ at both stripe boundaries $x=\pm a/2$ one recovers
the square-root behavior given by Eq.~(\ref{e:step_partcompcompvdw}).
In this limit the presence of the second, distant boundary of the
stripe gives rise to corrections to this square-root behavior
of the interfacial profile:
\begin{equation}\label{e:stripe_dwallcorr}
l(x;a\to\infty) = \left(\frac{8a_2^+}{\sigma_{lg}}\right)^{1/4}\sqrt{x}
\left[1-\frac{x}{2a}-\frac{x^2}{8a^2}+{\mathcal O}((x/a)^3)\right].
\end{equation}
The expression in Eq.~(\ref{e:stripe_dwallcorr}) corresponds to a
shifted coordinate system in which the two
boundaries of the stripe are located at $x=0$ and $x=a\to\infty$.

\subsection{Retardation}
\label{s:retardation}

Due to retardation
for large separations $r$ dispersion interaction potentials decay
as $r^{-7}$ rather than $r^{-6}$ \cite{israelachvili}. These retardation
effects become important for thick wetting layers. Therefore they have to
be taken into account in the discussion of the VDWT and of
the film morphology for the crossover between partially and completely
wetted substrate parts. For homogeneous and planar substrates in the
retarded regime the substrate potential decays as
$V(z\gg\sigma_f) = -v_4\,z^{-4}-v_5\,z^{-5} + {\mathcal O}(z^{-6})$.
If the interaction potential of the fluid particles is also retarded
one has $t(z\gg\sigma_f) = -s_4z^{-4}-s_5\,z^{-5} + {\mathcal
  O}(z^{-6})$ so that the effective interface potential turns into
\begin{equation}
\omega(l \gg \sigma_f) = \frac{b_3}{l^3}+\frac{b_4}{l^4} +
  {\mathcal O}(l^{-5})
\end{equation}
with $b_3 = \frac{\Delta\rho}{3}(v_4-s_4\rho_l)$
and $b_4 = \frac{\Delta\rho}{4}(v_5-(s_5+4s_4d_w)\rho_l)$.
Thus for critical wetting $l(T\nearrow T_w) = -4\,b_4/3\,b_3$
with $b_3$ changing sign at $T=T_w$ and $l(\Delta\mu \searrow 0) = 
(3\,b_3/\Delta\Omega_b)^{1/4} \sim (\Delta\mu)^{-1/4}$ along a complete
wetting isotherm.

The substrate potential of a SCS,
$V(x,z) = V_{att}(x,z)+V_{rep}(x,z)$ with $V_{att}(x,z)$ in accordance
with Eq.~(\ref{e:app_step_potattret}), approaches
\begin{equation}\label{e:step_spotretasym}
V_{att}(x\to\pm\infty,z) = -\frac{v_4^{\pm}}{z^4} +
\frac{v_4^+-v_4^-}{2\,x^4} + {\mathcal O}(x^{-5})
\end{equation}
for large $|x|$. Equation~(\ref{e:step_spotretasym}) implies
$\delta l_{\pm}(x) = l(x\to\pm\infty) - l_{\pm} \sim x^{-4}$
for the VDWT if the substrate $w_{\pm}$ undergoes a critical or
complete wetting transition. Analogously, for the LCS and the CST
one finds $\delta l_{\pm}(x) = l(x\to\pm\infty) - l_{\pm} \sim x^{-5}$.

For $\zeta=7$ Eqs.~(\ref{e:step_partcompcomp}) and
(\ref{e:step_partcompcrit}) imply
\begin{equation}
l(x\to\infty) =
\left(\frac{25\,b_3^+}{2\,\sigma_{lg}}\right)^{1/5}\,x^{2/5}
\end{equation}
in the case of complete wetting of $w_+$ and partial wetting of $w_-$
(i.e., $T_w^+ < T_0 < T_w^-$, $\Delta\mu=0$) and
\begin{equation}
l(x\to\infty) =
\left(\frac{18\,b_4^+}{\sigma_{lg}}\right)^{1/6}\,x^{1/3}
\end{equation}
at critical wetting (i.e., $T_w^+ = T_0 < T_w^-$, $\Delta\mu=0$).

\section{Numerical analysis of the interface morphology}
\label{s:numerical}

Although several properties of the liquid-vapor interfaces can be
determined analytically, even within the sharp-kink approximation the
behavior of $l(x)$ over the whole range of $x$ values
and for arbitrary values of $T$ and $\mu$ can only be obtained
numerically. Within the local theory for the SCS and the LCS we solve
the ELE given by Eq.~(\ref{e:step_elelocal}) and for the CST
Eq.~(\ref{e:stripe_eleloc}). The corresponding analysis within the
nonlocal theory requires to minimize numerically the functional
expression $\tilde{\omega}[l(x)]$ (see Eq.~(\ref{e:app_step_taudep}))
instead of solving the corresponding ELE. The reason is that as
pointed out in Ref.~\cite{bauer1} the numerical procedure
for solving this nonlocal ELE leads to severe difficulties. In
Ref.~\cite{bauer1} the
numerical techniques for calculating liquid-vapor 
interface profiles and a comparison of the results as obtained within
the nonlocal and the 
local theory are described in detail. It turns out that in general these
differences are very small. Therefore in the following we focus on the
numerical analysis of the local theory, keeping in mind that these
results are practically indistinguishable from those following from
the nonlocal theory although, in principle, the latter is the more
reliable one. 

\subsection{Chemical step}

Figure~\ref{f:step_complete} displays profiles of liquid-vapor interfaces for a
substrate with a chemical step undergoing complete wetting of both $w_+$ and
$w_-$. The parameters of the interactions are chosen such that both substrates
exhibit critical wetting transitions, $w_+$ at $T_w^* = k_BT_w/\epsilon_f = 1.0$
and $w_-$ at $T_w^* = 1.2$. The profiles of the 
liquid-vapor interfaces are calculated for $T^* = 1.3 > T_w^{\pm*}$ and
different values of $\Delta\mu$. In the limit $\Delta\mu\to 0$ both asymptotic
film thicknesses $l_{\pm}$ diverge according to $l_{\pm} \sim
(\Delta\mu)^{-1/3}$. The interface profile is extremely broad; we note
that the scales of the two axes differ by about three orders of
magnitude.

The crossover of $l(x)$ from $l_-$ to $l_+$ is governed by two
characteristic length scales, one for each substrate half $w_+$ and
$w_-$. Whereas along the isotherm the position $|x_0|\lesssim\sigma_f$
of the turning point remains practically unchanged, the two points
$x_+$ and $x_-$ where the profile $l(x)$ deviates 
from its asymptotes by $10\%$ of $|l_+-l_-|$ diverge according to a power
law. The characteristic length scale of the lateral variation of the profile
is governed by the lateral interfacial height-height correlation length
$\xi_{\parallel}$ given by Eq.~(\ref{e:step_corrlencomp}).
Figure~\ref{f:step_intwidthcomp} demonstrates that $x_{\pm}$ and
$\xi_{\parallel,comp}^{\pm}$ are proportional to each other.
                                
Figure~\ref{f:step_critical} shows liquid-vapor interface profiles
calculated for a SCS which exhibits a critical wetting transition at $T_w^* = 
1.0$ for both substrates $w_-$ and $w_+$. The interface profiles have
been obtained for different $T$ along the thermodynamic 
path $(T\to T_w, \Delta\mu=0)$. As before the position of the turning
points remains fixed at $|x_0|\lesssim\sigma_f$, the values of $x_+$
and $x_-$ diverge for $T\to T_w$, i.e., $l_{\pm}\to\infty$. Here the
divergence of $x_{\pm}$ is determined by $\xi_{\parallel,crit}^{\pm}$ given by
Eq.~(\ref{e:step_corrlencrit}). From Fig.~\ref{f:step_intwidthcrit} it
can be seen that also in the case of a critical
wetting transition $x_{\pm}$ and $\xi_{\parallel,crit}^{\pm}$ are
proportional to each other.

The behavior of $l(|x|\to\infty)$ is ultimately determined by the 
VDWT given in Eqs.~(\ref{e:step_vdwtcrit}) and
(\ref{e:step_vdwtcomp}). However, it turns out that these power-law
tails in $l(x)$ become relevant only for such large values of $|x|$ where 
in fact $\delta l(x)$ is smaller than the numerical error. Therefore
the analytically known VDWT are numerically not accessible. 

As demonstrated in Ref.~\cite{bauer1} the more general nonlocal theory
can be approximated by the corresponding local theory without losing
significant quantitative accuracy. Since the slope $dl/dx$ is small
one can go even a step further and replace within the local theory the
expression $\sqrt{1+(dl/dx)^2}-1$ by $(dl/dx)^2/2$ to be used in the
local functional $\tilde{\omega}_{loc}[l(x)]$ (see
Eqs.~(\ref{e:app_step_taudep}) and (\ref{e:app_locfunc})). This
gives rise to the ELE of a square-gradient theory:
\begin{eqnarray}\label{e:step_sqgrad_ele}
\sigma_{lg}l\,''(x)
& = & \Delta\Omega_b-\Delta\rho[\rho_l\,t(l(x)-d_w^+)-V(x,l(x))]
 \nonumber\\ 
& & + \Delta\rho\rho_l\left\{\bar{t}(x,l(x)-d_w^+) -
  \bar{t}(x,l(x)-d_w^-)\right\}
\end{eqnarray}
which is the analogue of a one-dimensional classical mechanical equation of
motion in a time-dependent external potential. Moreover, as an
additional approximation the actual substrate potential $V(x,z)$ with
a smooth lateral variation of $V_{att}$ (Eqs.~(\ref{e:app_step_potatt})
and (\ref{e:app_step_potrep})) can be replaced by the steplike
potential $V_{\infty}(x,z)=\Theta(-x)V_-(z)+\Theta(x)V_+(z)$. As it
turns out both approximations have only negligible effects on the
numerical results for the interface profiles (see Figs.~\ref{f:step_complete}
and \ref{f:step_critical}) because for thick wetting films the width
of the profile $l(x)$ is much larger than the scale of the lateral
variation of $V(x,z)$.

\begin{figure}
\begin{center}
\epsfig{file=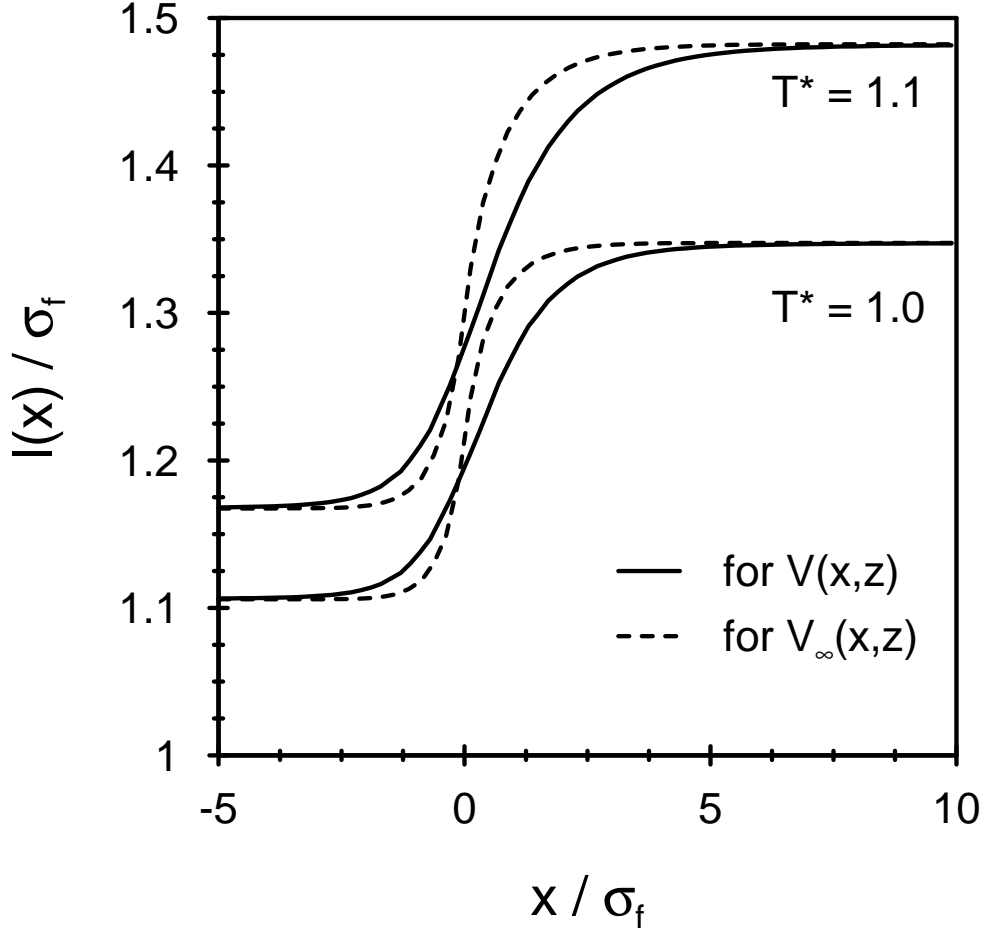, width=13cm, bbllx=70,
  bblly=320, bburx=520, bbury=780}
\end{center}
\caption{\label{f:step_1storder}
Liquid-vapor interface profiles across a SCS with the parameters chosen such that 
both substrates individually exhibit first-order wetting transitions,
the substrate $w_-$ at $T_w^* \approx 1.314$ (with $d_w^-=\sigma_f,
u_3^-=2.513\epsilon_f\sigma_f^3$,
$u_4^-=u_{4,x}^-=3.770\epsilon_f\sigma_f^4$, and
$u_9^-=0.335\epsilon_f\sigma_f^9$) and the substrate $w_+$ at $T_w^*
\approx 1.102$ (with $d_w^+=1.05\sigma_f$,
$u_3^+=3.710\epsilon_f\sigma_f^3$,
$u_4^+=u_{4,x}^+=5.566\epsilon_f\sigma_f^4$, and
$u_9^+=0.876\epsilon_f\sigma_f^9$). The system is at two-phase coexistence
$\Delta\mu=0$. The full and dashed lines correspond to the
full smooth substrate potential $V(x,z)$ and its steplike approximation
$V_{\infty}(x,z)$, respectively. In both cases the local ELE
(Eq.~(\ref{e:step_elelocal})) and its square-gradient
approximation (Eq.~(\ref{e:step_sqgrad_ele})) yield practically the
same results.}
\end{figure}

Figure~\ref{f:step_1storder} shows liquid-vapor interface profiles for a
SCS whose parameters are chosen such that first-order
wetting transitions occur at $T_w^* \approx 1.102$ and $T_w^* \approx 1.314$ on
$w_+$ and $w_-$, respectively. Here the liquidlike
layers are very thin. Therefore also the width of the profile is
rather small as compared to a system undergoing a critical or complete
wetting transition. In this case the replacement of the smooth substrate
potential $V(x,z)$ by the steplike potential $V_{\infty}(x,z)$ yields a
detectable difference but without changing the qualitative behavior.
Although the applicability of the present sharp-kink approximation for
such thin films certainly deserves a caveat we conclude
that only in the case of a substrate potential, which gives rise
to first-order wetting transitions, the transition region of the
interface profile on structured substrates is confined to a region of 
the order of $\sigma_f$ around the heterogeneity.

\subsection{Chemical step within a surface layer}

The previous subsection demonstrates that the properties of liquidlike
wetting layers on a SCS heterogeneity are determined to a large extent 
by the asymptotic substrate potentials $V_{\pm}(z)$ of the
semi-infinite substrate halves $x\lessgtr0$. This is also the case for
the LCS. The wetting properties of the substrate halves $x\lessgtr0$
give rise to the asymptotic thickness 
of the wetting layer $l(x\to\pm\infty)=l_{\pm}$ as discussed in
Subsec.~\ref{s:hom_dft} and $l(x)$ smoothly and monotonously
interpolates between them. Figure~\ref{f:2sl_diffd} displays a typical
example for the influence of the thickness of the surface layer $n$ on
the profiles $l(x)$. This system exhibits a critical wetting transition at
$T_w^{+*}=T_w^{-*}=T_w^*=1.2$; we choose $T^*=1.1$ and $\Delta\mu^* =
1.5\cdot10^{-6}$. Without an overlayer (i.e., the number of
inhomogeneous surface layers $n=0$) the system corresponds to the
homogeneous and planar substrate $w_H$. As $n$ is increased there is a
crossover of $l(x)$ to the profile of a SCS. The typical profile widths are
-- as for the SCS -- given by Eqs.~(\ref{e:step_corrlencomp})
and (\ref{e:step_corrlencrit}) if one of the substrate halves exhibits a
complete or a critical wetting transition, respectively. In the case
of a first-order wetting transition, i.e., for thin liquidlike layers
the deviation of $l(x)$ from its asymptotes is confined to a region of
the order of $\sigma_f$ around $x=0$.

From the analytical results for the VDWT one expects that the profile
$l(x)$ for a LCS approaches its asymptotes $l_{\pm}$ faster  
than in the case of a SCS, i.e., via $\delta l(|x|\to\infty) \sim x^{-4}$
(see Eqs.~(\ref{e:2sl_vdwt})--(\ref{e:2sl_vdwtcomp})) rather than
$x^{-3}$ (Eqs.~(\ref{e:step_vdwtcrit}) and (\ref{e:step_vdwtcomp})). However,
since as observed above the VDWT become relevant only for very large
values of $|x|$ without affecting the main deviation of $l(x)$ from
$l_{\pm}$ around $x=0$, this more rapid decay is 
analytically accessible but does not play an important role for the
overall behavior of $l(x)$.

The divergence of the function $l(x)$ for $x\to\infty$ in the case
that the substrate half $x>0$ is completely and the other substrate
half $x<0$ is only partially wet is given by
Eqs.~(\ref{e:step_partcompcomp}) and (\ref{e:step_partcompcrit}) for
$T>T_w^+$ and $T=T_w^+$, respectively. This behavior is shown in
Figs.~\ref{f:2sl_partcomp} and \ref{f:2sl_partcrit}. The asymptotic
behavior for $x>0$ of the profiles for small $\Delta\mu$ and small
$T_w-T$, respectively, is in excellent agreement with the numerical
solutions of Eqs.~(\ref{e:step_deforcomp}) and (\ref{e:step_deforcrit}).

\subsection{Chemical stripe}

First we analyze a CST for which the homogeneous substrate $w$
corresponding to the embedding substrate undergoes a critical
wetting transition at $T_w^*=1.2$ whereas the homogeneous substrate
corresponding to the inserted slab $w_{st}$ exhibits a critical wetting
transition at $T_w^*=1.0$. We consider thermodynamic states
along the complete wetting isotherm $(T^*=1.1,\Delta\mu)$ so that at
coexistence the substrate as a whole is only partially wet. The complete wetting
transition on $w_{st}$ is suppressed by the finite lateral extension of
$w_{st}$.

Figure~\ref{f:stripe_complete} displays liquid-vapor interface
profiles for different values of $\Delta\mu$. The width of the
transition regions at $x=\pm a/2$ increases as $\Delta\mu$
becomes smaller. If this width is small compared with the width $a$
of the stripe region the profile $l(x)$ attains the equilibrium film
thickness $l_+$ on the stripe area. Therefore, if $a$ is large the 
profile for a CST is approximately composed of two interfacial
profiles corresponding to two SCS located at
$x=-a/2$ and $x=+a/2$. As the stripe width $a$ is decreased for
fixed values of $T$ and $\Delta\mu$ the region where $l(x)$ attains
the value $l_+$ decreases and ultimately vanishes as $a$ becomes
smaller than the width of the transition region. This behavior is shown in
Fig.~\ref{f:stripe_diffwidth}. 

Figure~\ref{f:stripe_exchanged} presents interface profiles for a CST
with the chemical species of the slab and the surrounding substrate
exchanged. Therefore the substrate exhibits a critical wetting
transition at $T_w^*=1.0$ and it is already completely wet at
coexistence as the
temperature is raised towards $T^*=1.2$. The latter is the transition
temperature of the critical wetting transition on the homogeneous
substrate corresponding to $w_{st}$. If we choose $T^*=1.1$ and decrease 
$\Delta\mu$ the whole substrate undergoes a complete wetting
transition and the ``dent'' induced by the existence of the
stripe-shaped heterogeneity is smeared out.

\begin{figure}
\begin{center}
\epsfig{file=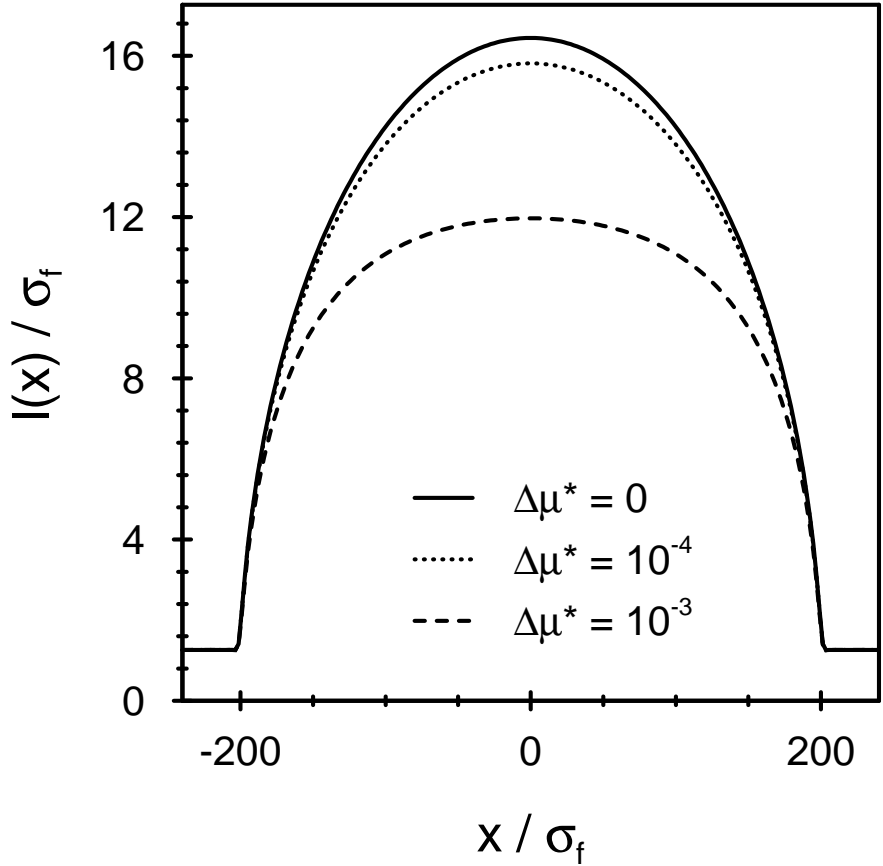, width=12cm, bbllx=80,
  bblly=320, bburx=520, bbury=775}
\end{center}
\caption{\label{f:stripe_confinement}
Liquid-vapor interface profiles on a CST which undergoes a first-order
wetting transition for a choice of temperature such that the
substrate corresponding to $w_{st}$ is completely wet
(``hydrophilic'') whereas the embedding substrate is only 
partially wet (``less hydrophilic''). The parameters are chosen as in
Fig.~\ref{f:step_1storder}. Therefore the outer region undergoes a
first-order wetting transition at $T_w^{-*}\approx1.314$ whereas the
homogeneous substrate corresponding to $w_{st}$ exhibits a
first-order wetting transition at $T_w^{+*}\approx1.102$. The temperature
is taken as $T^*=1.2$ so that $T_w^+<T<T_w^-$. The width of the stripe is
$a=400\sigma_f$ and the chemical potential is varied. Since the
equilibrium film thickness $l_- = l(|x|\to\infty)$ is of the order of
$\sigma_f$ the liquidlike ``channel'' is confined to the
``hydrophilic'' stripe region without leaking. The shape of the
liquidlike channel for 
$\Delta\mu=0$ is in good agreement with the semi-elliptic shape given by
Eq.~(\ref{e:stripe_ellipse}). If the stripe 
width $a$ is increased, $l(x=0)$ and $l(x=0)-l_-$ increase as $a^{1/2}$
and the excess coverage supported by the stripe diverges as
$\Gamma_{ex}\sim a^{3/2}$.}
\end{figure}

The above analysis demonstrates that the liquid leaks out of the
chemical stripe if on the embedding material the fluid is close to a
complete or critical wetting
transition. Figure~\ref{f:stripe_confinement} shows that this leaking
is absent if on the embedding material the fluid can form only a thin
wetting film. This situation prevails if the outer material leads to a
first-order wetting transition at $T_w^-$ and the temperature is
chosen such that $T<T_w^-$. Figure~\ref{f:stripe_confinement}
corresponds to a CST for which the slab as a homogeneous substrate
undergoes a first-order wetting transition, too, at $T_w^+$ with
$T_w^+<T<T_w^-$. The confinement of the liquid to the stripe is
achieved both off and at two-phase coexistence. In view of practical
applications of chemically structured surfaces in the context of
microfluidics this tells for which materials and for which
thermodynamic states a good performance without leakage can be
expected.

In the strict sense of thermal equilibrium so far only a few
solid-fluid systems are known to exhibit a true first-order wetting
phase transition at coexistence (e.g., He$^4$ on
Cs~\cite{chengetal,hallock}); up to now critical wetting has been 
observed only on fluid substrates (e.g., alkanes on aqueous solutions
of salt or glucose~\cite{ragiletal,pfohlriegler}). The generic case is
that fluids wet substrates completely above the triple
point~\cite{sdreview,beysens}. Although our theoretical framework is based on
equilibrium statistical mechanics, the present model calculations for
the morphology of wetting films can be applied directly also to
nonvolatile liquids as long as their interaction with solid substrates
can be modelled by appropriate effective interface potential
$\omega_{\pm}(l)$. Alternatively, these potentials can be inferred
empirically from suitable experiments (see, e.g.,
Ref.~\cite{kimetal,herminghausetal}). In this sense our conclusions are
valid also for nonvolatile liquids adsorbed on ``hydrophilic'' or
``hydrophobic'' parts of an inhomogeneous substrate, for which a wide
range of wettability characteristics, beyond the strict equilibrium
conditions, can be arranged.

\begin{figure}
\begin{center}
\epsfig{file=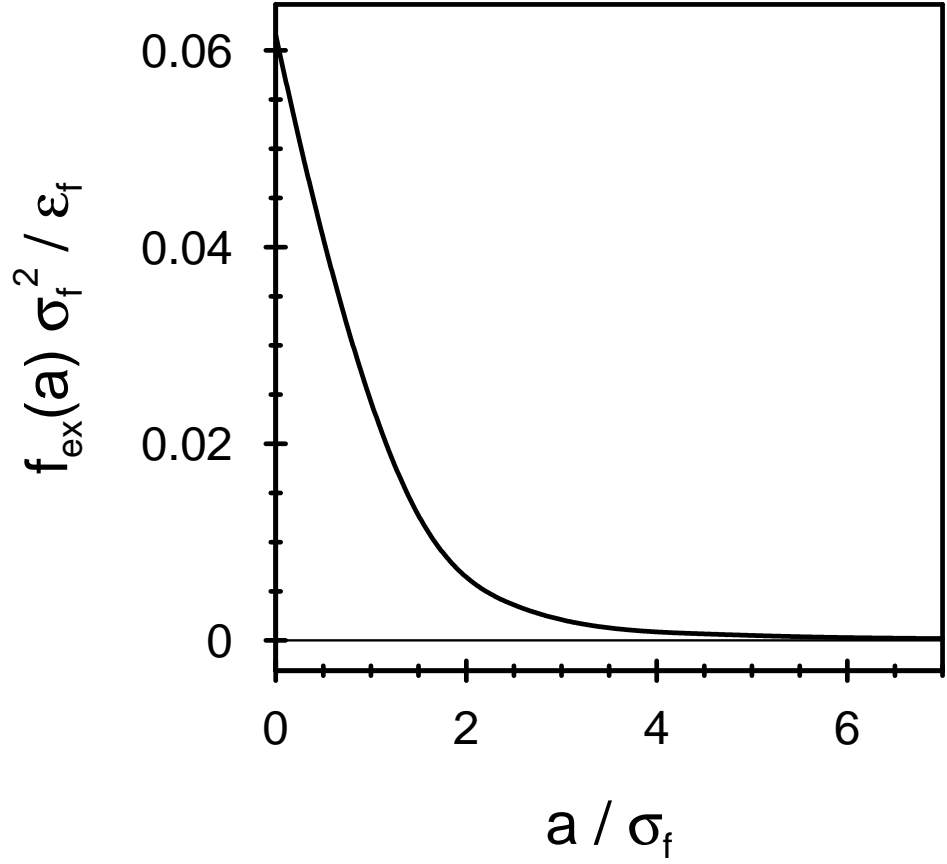, width=13cm, bbllx=45, bblly=330,
  bburx=460, bbury=715}
\end{center}
\caption{\label{f:stripe_force}
Excess contribution $f_{ex}(a)=-\partial\delta\tau(a)/\partial a$ to
the force per unit length which acts on a chemical stripe. The system is the same as
in Fig.~\ref{f:stripe_confinement} but with $T^*=1.0$ and at two-phase
coexistence $\Delta\mu=0$. The total force is $f(a)=f_0+f_{ex}(a)$,
i.e., it is the sum of $f_{ex}(a)$ and the constant contribution
$f_0=\sigma_{w_-g}-\sigma_{w_+g}$ which in the present example is
$f_0\approx0.137\epsilon_f/\sigma_f^2$. In the limit $a\to\infty$
$f_{ex}(a)$ decays as $a^{-3}$. Both $f_0$ and $f_{ex}$ are positive,
leading to a dilation of the stripe.}
\end{figure}

The line tension $\tau_{CST}(a)$ associated with a chemical stripe can
be written as
\begin{equation}\label{e:stripe_force1}
\tau_{CST}(a) = a(\sigma_{w_+g}-\sigma_{w_-g}) +
2\tau_{SCS} + \delta\tau(a)
\end{equation}
where $\sigma_{w_{\pm}g}$ is the
surface tension of the homogeneous ``$+$'' or ``$-$'' substrate in contact
with the vapor, $\tau_{SCS}$ is the line tension associated with a
single SCS, and $\delta\tau(a)$ with, as it turns out,
$\delta\tau(a\to\infty)\sim a^{-2}$ is the effective interaction between the
two line structures a distance $a$ apart from each other.
$\tau_{CST}(a)$ generates a lateral force per unit length 
\begin{equation}\label{e:stripe_force2}
f(a)\equiv-\frac{\partial\tau_{CST}(a)}{\partial a}
\end{equation}
acting on the stripe and leading to a compression or dilation which is
balanced by the elastic forces of the substrate. $f(a)$ is defined such
that a positive sign corresponds to a dilation of the stripe. Our
numerical results show that for large stripe widths and independent of
the wetting characteristics of the materials under consideration 
$f(a)$ is practically constant with 
$f(a) = \sigma_{w_-g}-\sigma_{w_+g} +
{\mathcal O}(a^{-3})$, i.e., in Eq.~(\ref{e:stripe_force1}) for
$a\gtrsim4\sigma_f$ the first contribution is the dominating one
(see Fig.~\ref{f:stripe_force}). For solid substrates the force is expected to have
practically no effect because the compressibility of the substrate is
very small. However, for fluid substrates such as fluids covered with
Langmuir-Blodgett films the force can have a significant
effect. Therefore we propose to test this effective force between line 
structures by adsorbing liquidlike wetting films on a liquid
substrate decorated with a Langmuir-Blodgett film which contains a
stripe of different material. The adsorbed wetting films will cause
a change in the stripe width depending on the compressibility of the
Langmuir-Blodgett films. Moreover, such experimental arrangements
would facilitate to probe our predictions concerning critical
wetting transitions, as reported in Refs.~\cite{ragiletal} and
\cite{pfohlriegler}, on chemically structured substrates.

\section{General model for chemically heterogeneous substrates}
\label{s:general}
\newcommand{\vecr}{{\mathbf r}}
\newcommand{\vecrp}{{\mathbf r}_{\parallel}}

\begin{figure}
\begin{center}
\epsfig{file=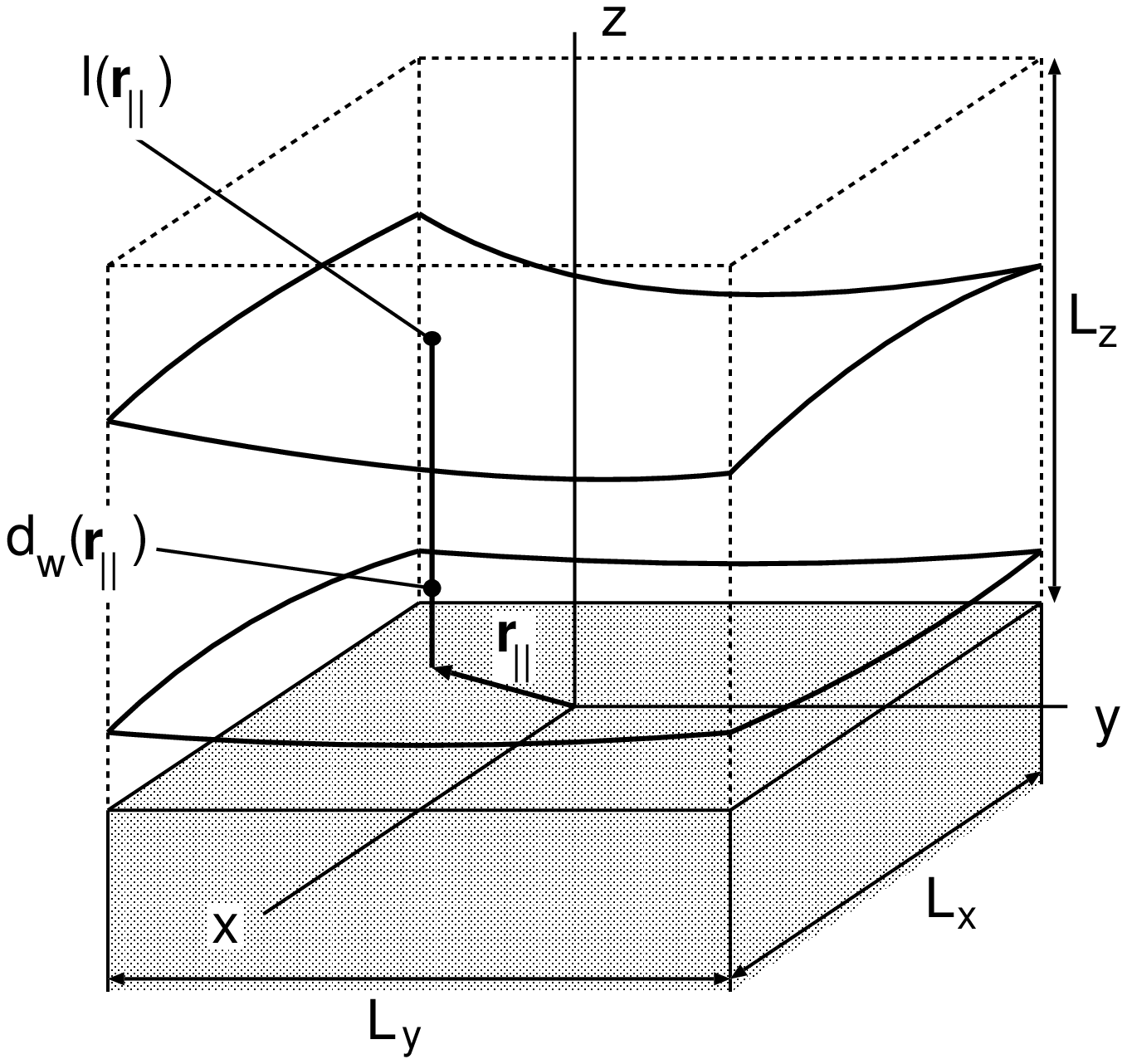, width=13cm}
\end{center}
\caption{\label{f:generalinh}
Sketch of a fluid film on a general chemical inhomogeneity. The
potential $V(\vecrp,z)$ of the arbitrarily structured substrate (with
$\vecrp=(x,y)$) gives rise to a laterally varying exclusion length
$d_w(\vecrp)$ and a liquid-vapor interface shape described by
$l(\vecrp)$. The surface $(\vecrp,z=l(\vecrp))$ separates the
liquidlike wetting layer (below) from the bulk vapor phase
(above). The system is truncated at $x=\pm L_x/2$, 
$y=\pm L_y/2$, and $z=L_z$ in order to facilitate the proper
thermodynamic limit. (This truncation gives rise to artificial surface
and line tensions generated by these boundaries.) The chemical
inhomogeneity of the substrate is not indicated.}
\end{figure}

It is possible to generalize the methods developed within the previous
sections to substrates with arbitrary heterogeneities. The only
required input is the substrate
potential which in general is a function $V(\vecr)$ of all coordinates
$\vecr=(\vecrp,z)=(x,y,z)$ and approximately follows from the
summation of pair potentials. The substrate potential $V(\vecr)$,
together with the fluid-fluid interaction, gives rise to an interface
profile $l(\vecrp)$ which in general depends on both lateral
coordinates $\vecrp=(x,y)$ (see
Fig.~\ref{f:generalinh}). We assume that the substrate is flat and
located in the half space $w = \{\vecr\in{\mathbb R}^3|z\leq 0\}$.
We take into account the excluded volume at the substrate 
surface by introducing a spatially varying
$d_w(\vecrp)$. The insertion of the sharp-kink density profile 
\begin{equation}\label{e:generalsharpkink}
\hat{\rho}(\vecr) = \Theta(z-d_w(\vecrp))[\Theta(l(\vecrp)-z)\rho_l +
\Theta(z-l(\vecrp))\rho_g]
\end{equation}
into Eq.~(\ref{e:ldafunctional}) and the decomposition of
$\Omega[\hat{\rho}]$ into bulk and subdominant contributions yields
\begin{equation}\label{e:generaldecomposition}
\Omega[\hat{\rho}(\vecr)] = \Lambda\Omega_b(\rho_g;T,\mu) +
\Omega_s([l(\vecrp)];T,\mu;[\tilde{w}],[V]).
\end{equation}
Concerning the expression for $\Omega_s$ we consider the thermodynamic
limit and omit artificial contributions which stem from the truncation
of the system. $\Lambda=L_xL_yL_z$ is the volume filled with fluid particles
and $\Omega_b$ is the free energy density of the bulk vapor phase
(see Eq.~(\ref{e:bfedensity})). The subdominant contribution in
Eq.~(\ref{e:generaldecomposition}) reads 
\begin{eqnarray}\label{e:generalsurfacecont}
\Omega_s[l(\vecrp)] & = &
\int_A
d^2r_{\parallel}\,\{\Delta\Omega_b\,l(\vecrp) +
\sigma_{wl}(\vecrp;d_w(\vecrp))+ \Sigma_{lg}(\vecrp,[l(\vecrp)]) \nonumber\\ 
& & + \omega(\vecrp,l(\vecrp);d_w(\vecrp)) + {\mathcal
  C}(\vecrp,l(\vecrp);[d_w(\vecrp)])\}
\end{eqnarray}
where
\begin{equation}\label{e:generaldepwlstens}
\sigma_{wl}(\vecrp;d_w(\vecrp)) =
-\frac{\rho_l^2}{2}\int\limits_0^{\infty} dz\,t(z) 
+ \rho_l\int\limits_{d_w(\vecrp)}^{\infty} dz\,V(\vecr) -
d_w(\vecrp)\Omega_b^{(l)}
\end{equation}
can be interpreted as a local, spatially varying wall-liquid surface tension.
Equation~(\ref{e:generaldepwlstens}) is a generalization of
Eq.~(\ref{e:hom_wlstens}). 
$\Sigma_{lg}(\vecrp,l(\vecrp))$ is the surface free energy density
containing the cost in free energy for deforming the liquid-vapor
interface:
\begin{equation}\label{e:generallgfednloc}
\Sigma_{lg}^{(nloc)}(\vecrp,[l(\vecrp)]) = \sigma_{lg}-\frac{(\Delta\rho)^2}{2}
\int_A d^2r_{\parallel}\,'\, 
\int\limits_0^{\infty} dz\int\limits_0^{l(\vecrp)-l(\vecrp\,')}dz' \,
\tilde{w}(|\vecrp-\vecrp\,'|,|z-z'|)
\end{equation}
within the nonlocal and
\begin{equation}\label{e:generallgfedloc}
\Sigma_{lg}^{(loc)}(\vecrp,[l(\vecrp)]) =
\sigma_{lg}\sqrt{1+(\nabla_{\parallel}l(\vecrp))^2} 
\end{equation}
within the local theory with $\nabla_{\parallel} \equiv
(\partial_x,\partial_y)$. The latter expression is the leading term
of the gradient expansion of $\Sigma_{lg}^{(nloc)}$. In
Eqs.~(\ref{e:generallgfednloc}) and (\ref{e:generallgfedloc})
$\Sigma_{lg}$ does not depend on the absolute value of $l(\vecrp)$
measured from the substrate surface but only on the relative
differences $l(\vecrp)-l(\vecrp')$ so that $\sigma_{lg}$ is
independent of $l(\vecrp)$. Such additional dependences are brought
about by replacing the sharp-kink density profile in
Eq.~(\ref{e:generalsharpkink}) by 
a smooth one whose tails are cut off by the surface (see, e.g.,
Refs.~\cite{fisherjin} and \cite{parryevans}). As a generalization of
Eq.~(\ref{e:app_localeffintpot}) 
\begin{equation}\label{e:localeffintpot}
\omega(\vecrp,l(\vecrp);d_w(\vecrp)) =
\Delta\rho\,\rho_l\int\limits_{l(\vecrp)-d_w(\vecrp)}^{\infty} 
dz\,t(z) - \Delta\rho\int\limits_{l(\vecrp)}^{\infty}dz\,V(\vecr)
\end{equation}
is the ``local'' effective interface potential for the effective
interaction between the substrate surface and the liquid-vapor interface.
Finally,
\begin{eqnarray}
\lefteqn{ {\mathcal C}(\vecrp,l(\vecrp);[d_w(\vecrp)]) = 
- \frac{\rho_l^2}{2}\int\limits_0^{\infty}dz\,t(z)
+\rho_l\,\Delta\rho\int\limits_{d_w(\vecrp)-l(\vecrp)}^{\infty}dz\,t(z) }
\nonumber\\
& & + \frac{\rho_l^2}{2}\int_A d^2r_{\parallel}\,'
\int\limits_{d_w(\vecrp)-d_w(\vecrp\,')}^{\infty}dz \int\limits_z^{\infty}dv
\,\tilde{w}(|\vecrp-\vecrp\,'|,|v|) \nonumber\\
& & - \rho_l\,\Delta\rho\int_A d^2r_{\parallel}\,'
\int\limits_{d_w(\vecrp\,')-l(\vecrp)}^{\infty}dz \int\limits_z^{\infty}dv
\,\tilde{w}(|\vecrp-\vecrp\,'|,|v|)
\end{eqnarray}
takes into account the effects due to the lateral variation of
$d_w(\vecrp)$. ${\mathcal C}$ vanishes for $d_w(\vecrp) \equiv d_w =
\mbox{const}$. 

If the substrate contains inhomogeneities with large linear extensions, i.e., the
substrate is translationally invariant along one of the lateral
directions, the subdominant contribution $\Omega_s$
decomposes into ``true'' surface and line contributions as discussed
above for the SCS, the LCS, and the CST. In the absence of such an
additional translational symmetry one cannot identify a genuine line
contribution.

The equilibrium liquid-vapor interface profile $\bar{l}(\vecrp)$
minimizes $\Omega_s$. Inserting $\bar{l}(\vecrp)$ into $\Omega_s$
renders an the equilibrium, laterally varying wall-vapor surface
tension $\sigma_{wg}(\vecrp)$ such that
\begin{equation}
\int_A d^2r_{\parallel}\,\sigma_{wg}(\vecrp) =
\min_{\{l(\vecrp)\}}\Omega_s[l(\vecrp)] = \Omega_s[\bar{l}(\vecrp)]
\end{equation}
where (see Eq.~(\ref{e:generalsurfacecont}))
\begin{eqnarray}\label{e:generalwallgastens}
\sigma_{wg}(\vecrp) & = & \Delta\Omega_b\,\bar{l}(\vecrp) +
\sigma_{wl}(\vecrp;d_w(\vecrp))+ \Sigma_{lg}(\vecrp,[\bar{l}(\vecrp)]) \nonumber\\ 
& & + \omega(\vecrp,\bar{l}(\vecrp);d_w(\vecrp)) + {\mathcal
  C}(\vecrp,\bar{l}(\vecrp);[d_w(\vecrp)]).
\end{eqnarray}
Within the local theory and for
$d_w(\vecrp) \equiv d_w$ Eq.~(\ref{e:generalwallgastens}) reduces to
\begin{equation}
\sigma_{wg}(\vecrp) = \Delta\Omega_b\,\bar{l}(\vecrp) + \sigma_{wl}(\vecrp) +
\sigma_{lg}\sqrt{1+(\nabla_{\parallel}\bar{l}(\vecrp))^2} +
\omega(\vecrp,\bar{l}(\vecrp))
\end{equation}
with the corresponding ELE
\begin{eqnarray}\label{e:generalele}
\lefteqn{ \sigma_{lg}\nabla\cdot
\left(\frac{\nabla\bar{l}(\vecrp)}{\sqrt{1+(\nabla\bar{l}(\vecrp))^2}}\right) 
= \Delta\Omega_b + \left.\frac{\partial\omega(\vecrp,l)}{\partial
    l}\right|_{l=\bar{l}(\vecrp)} } \nonumber\\
& = & \Delta\Omega_b - \Delta\rho\,\rho_l\,t(\bar{l}(\vecrp)-d_w) +
\Delta\rho\,V(\vecrp,\bar{l}(\vecrp)).
\end{eqnarray}
This type of equation is often used (e.g., in Ref.~\cite{swainlipowsky}) to
study the macroscopic properties of liquidlike wetting layers and
droplets on structured substrates. This section provides a microscopic
basis for the underlying concept of a local, spatially varying surface
tension as it is used in several studies (e.g., in
Refs.~\cite{lenzlipowsky,swainlipowsky,lipowskylenzswain}).

\section{Summary}
\label{s:summary}

\begin{figure}
\begin{center}
\epsfig{file=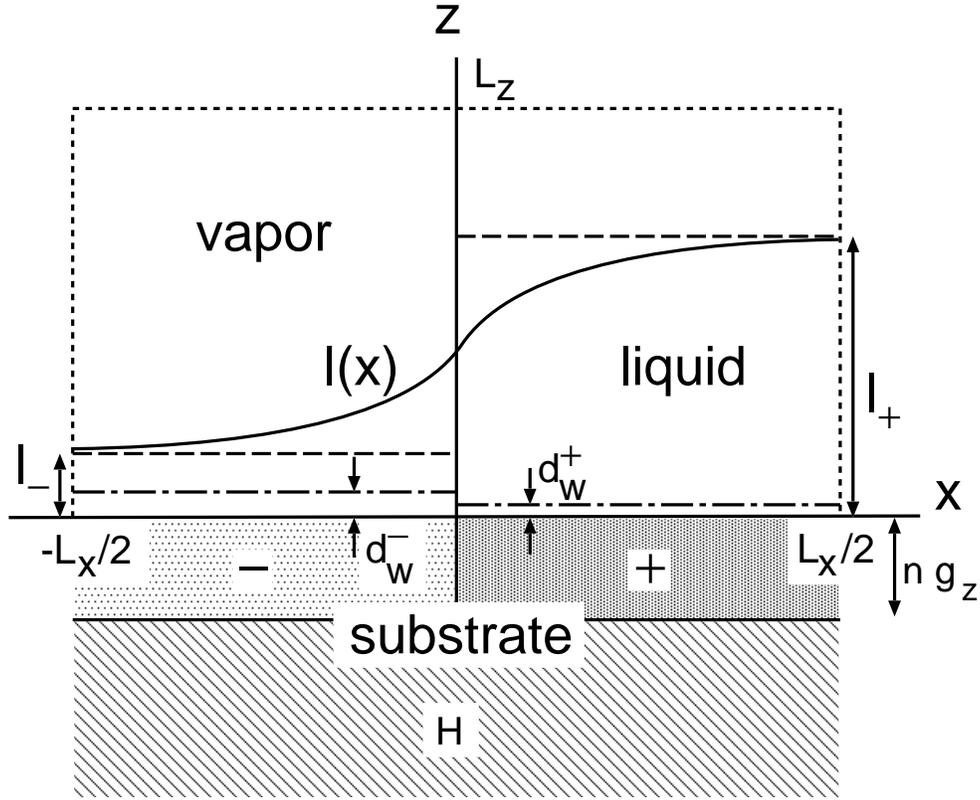, width=13cm}
\end{center}
\caption{\label{f:step_2sl_system}
Morphology $l(x)$ of a liquidlike film which covers a flat, heterogeneous
substrate (layer chemical step (LCS)). The substrate
surface is located at $z=0$. In general the substrate consists of a
homogeneous part (denoted as $H$) which is covered by two adjacent,
semi-infinite surface layers composed of different chemical species
(``$+$'' and ``$-$'') which form a sharp chemical boundary at $x=0$. The
surface layer consists of $n=d+1$ molecular 
monolayers with a lattice spacing $g_z$ in the $z$ direction. The
limit $n\to\infty$ corresponds to a substrate which exhibits a single
``simple chemical step'' (SCS). $l_{\mp} = l(x\to\mp\infty)$ are the
equilibrium film thicknesses of the corresponding substrates composed
of ``$H$'' plus a 
laterally homogeneous surface layer (``$+$'' or ``$-$''). We assume
that ``$H$'' covered by a ``$+$'' layer is the stronger substrate favoring
the adsorption of thicker liquidlike films, i.e., $l_+>l_-$. The two
substrate halves are characterized by different excluded volumes
$d_w^{\pm}$ (dash-dotted lines). The system is
translationally invariant in the $y$ direction and truncated at
$z=L_z$ and $x=\mp L_x/2$ in order to facilitate the proper
thermodynamic limit. The long-dashed line corresponds to the sharp-kink
approximation for the \emph{lateral} profile.}
\end{figure}

\begin{figure}
\begin{center}
\epsfig{file=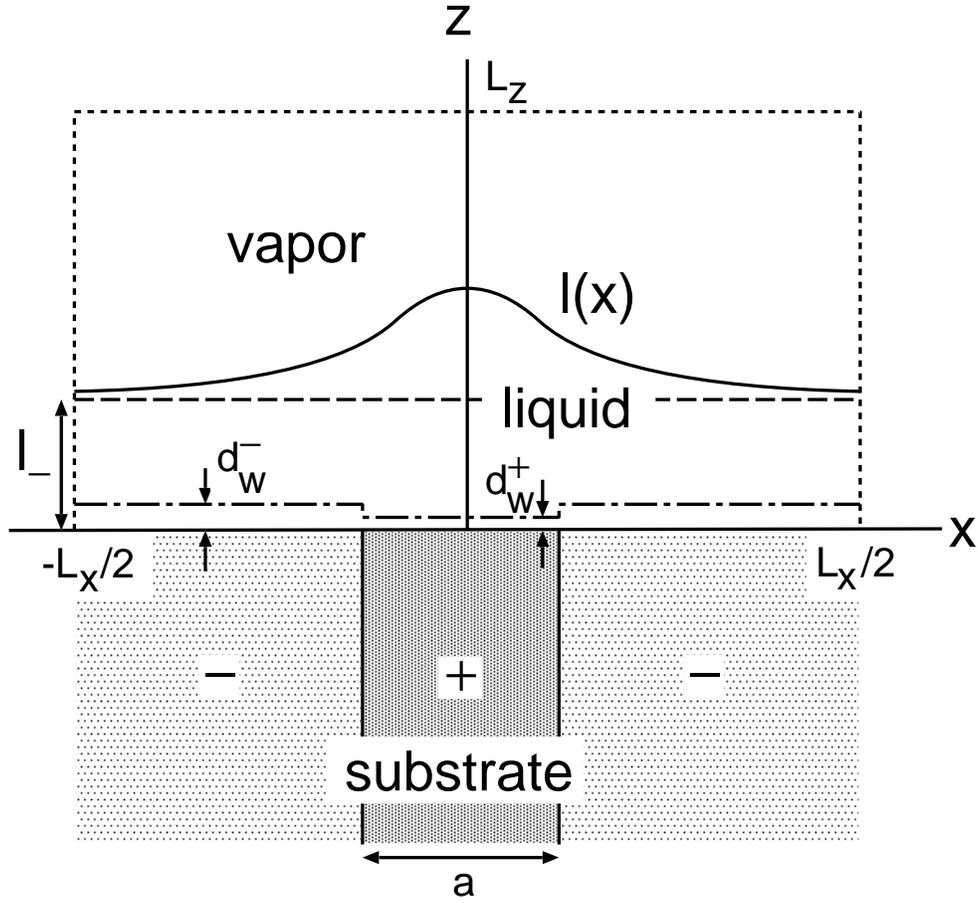, width=13cm}
\end{center}
\caption{\label{f:stripe_system}
Morphology of a liquidlike film which covers a flat substrate exhibiting
a ``chemical stripe'' (CST). A slab consisting of chemical species
denoted by ``$+$'' is immersed in a homogeneous ``$-$''
substrate. Thus in top view a chemical stripe with two sharp chemical 
steps is formed. The slab extends from $x=-a/2$ to $x=a/2$, i.e., the
stripe width is $a$. $l_- = l(|x|\to\infty)$ is the equilibrium thickness of the
liquidlike wetting film corresponding to the homogeneous substrate
composed of ``$-$'' particles. In contrast to the SCS and the LCS
(compare Fig.~\ref{f:step_2sl_system}), here both cases $l_+>l_-$ and
$l_+<l_-$ -- where $l_+$ corresponds to the film thickness on the
homogeneous ``$+$'' substrate -- have to be considered. The slab and
the embedding substrate exhibit different excluded volumes
$d_w^{\pm}$. The system is translationally invariant in the $y$
direction and truncated at $z=L_z$ and $x=\mp L_x/2$ in order to
facilitate the proper thermodynamic limit.}
\end{figure}

\begin{figure}
\begin{center}
\epsfig{file=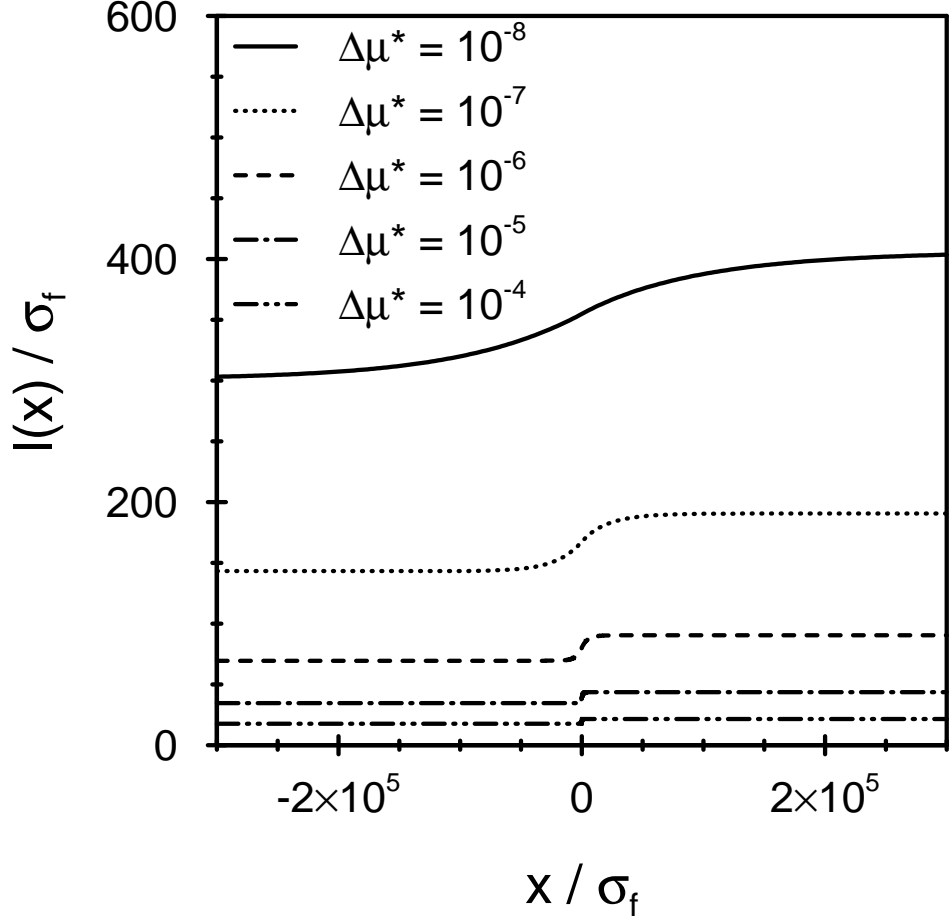, width=13cm, bbllx=65,
  bblly=320, bburx=520, bbury=775}
\end{center}
\caption{\label{f:step_complete}
Shape of the liquid-vapor interface profile on a SCS (see
Fig.~\ref{f:step_2sl_system}, with $n=\infty$) along an isotherm
$(T^*=1.3, \Delta\mu^*)$, i.e., for complete wetting of both substrates
$w_+$ and $w_-$. The parameters are chosen such that both substrates 
individually exhibit critical wetting transitions, $w_+$ at
$T_w^*=1.0$ and $w_-$ at $T_w^*=1.2$: $d_w^{\pm}=\sigma_f$,
$u_3^+=2.079\epsilon_f\sigma_f^3$,
$u_{4}^+=12.475\epsilon_f\sigma_f^4$, $u_{4,x}^+ = u_4^+g_x/g_z =
u_4^+$, and $u_9^+=0.277\epsilon_f\sigma_f^9$, whereas
$u_i^-=0.809u_i^+$. Therefore for the temperature chosen here
at two-phase coexistence both substrate halves are completely wet. 
The profiles shown here remain practically unchanged if the local ELE
in Eq.~(\ref{e:step_elelocal}) is approximated further by the
square-gradient expression in Eq.~(\ref{e:step_sqgrad_ele}) or if the
substrate potential $V(x,z)$ is replaced by the steplike substrate
potential $V_{\infty}(x,z) = V_+(z)\Theta(x)+V_-(z)\Theta(-x)$.}
\end{figure}

\begin{figure}
\begin{center}
\epsfig{file=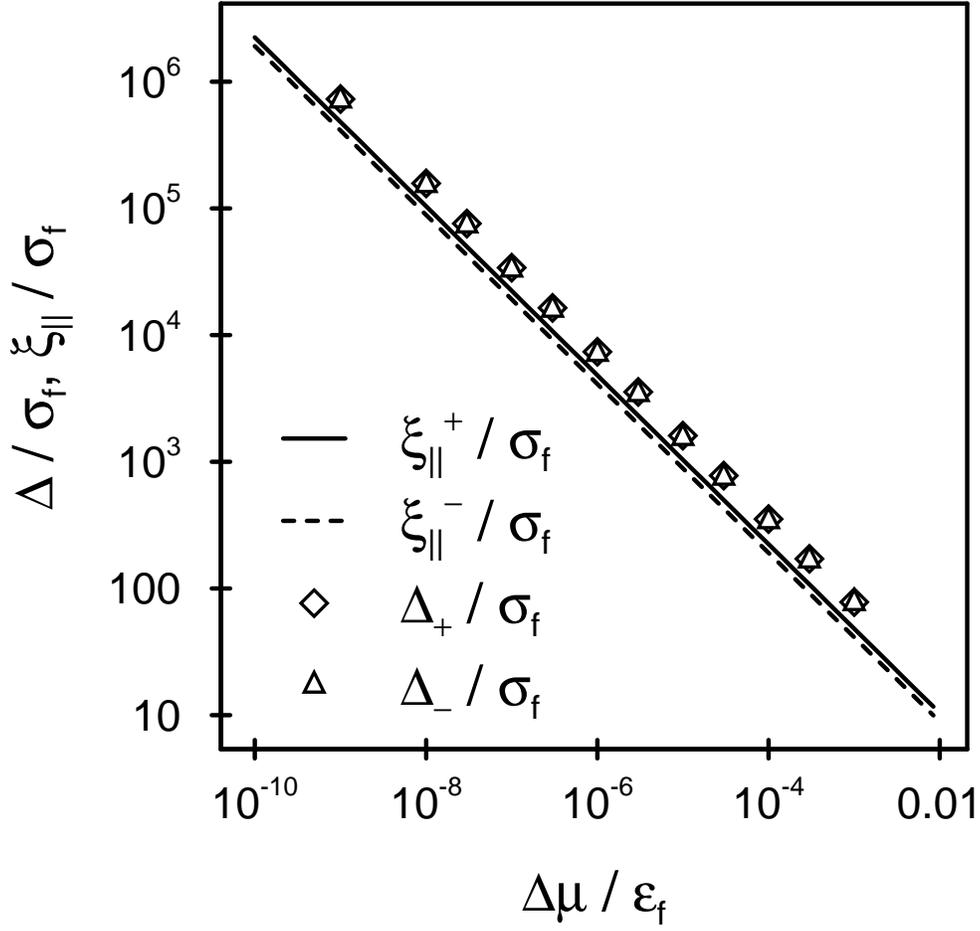, width=13cm, bbllx=70,
  bblly=315, bburx=515, bbury=770}
\end{center}
\caption{\label{f:step_intwidthcomp}
Widths $\Delta_{\pm}$ of the interface profiles ($\lozenge$, $\triangle$)
shown in Fig.~\ref{f:step_complete} compared with the lateral
height-height correlation lengths $\xi_{\parallel}^{\pm}$ determined
by $l_{\pm}$ (full and dashed line,
Eq.~(\ref{e:step_corrlencomp})). $\Delta_{\pm}$ is defined as that
value $|x|$ at which $l(x)$ starts to deviate from its respective
asymptote $l_{\pm}$ by $10\%$ of $|l_+-l_-|$ and measures the lateral
extension of the transition region of the interface profile. For
$\Delta\mu\to0$ both $\Delta_{\pm}$ and $\xi_{\parallel}^{\pm}$
diverge according to $(\Delta\mu)^{-2/3}$ as expected for
complete wetting transitions.}
\end{figure}

\begin{figure}
\begin{center}
\epsfig{file=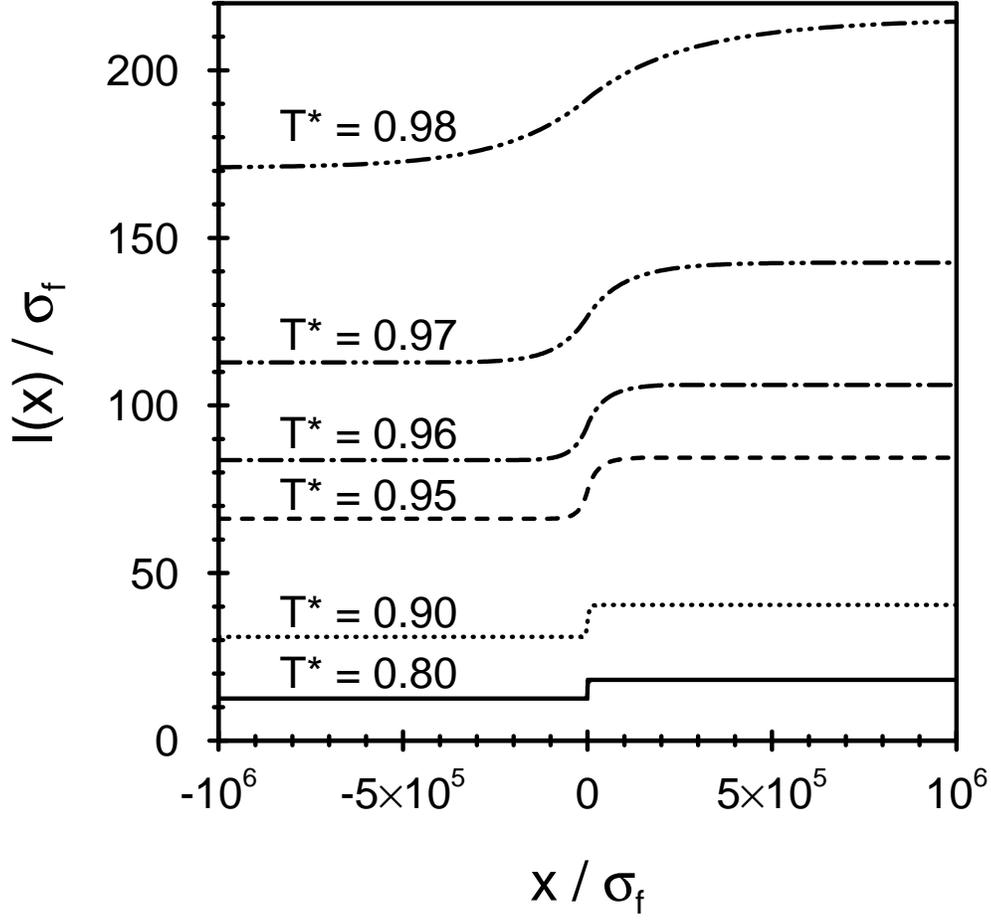, width=13cm, bbllx=70,
  bblly=330, bburx=520, bbury=775}
\end{center}
\caption{\label{f:step_critical}
Shape of the liquid-vapor interface profile of a liquidlike film across a
SCS on a thermodynamic path at coexistence $\Delta\mu=0$ with the
temperature $T^*$ increasing towards the wetting transition
temperature $T_w^*=1.0$ characteristic for \emph{both} substrates
$w_{\pm}$. The parameters are chosen such that the wetting transitions
are continuous: $u_i^{\pm}$ and $u_{i,x}^{\pm}$ are equal to
$u_i^+$ as in Fig.~\ref{f:step_complete} but with
$u_4^+=u_{4,x}^+=14.035\epsilon_f\sigma_f^4$ and $d_w^{\pm}$ are also
as in Fig.~\ref{f:step_complete}. Therefore $l_+(T) \neq
l_-(T)$ although $T_w^+=T_w^-$. Upon approaching the wetting transition
temperature both $l_+$ and $l_-$ diverge as $t_w^{-1}$ with
$t_w = (T_w-T)/T_w$. As in Fig.~\ref{f:step_complete} the results are
practically insensitive against using the square-gradient
approximation or using a steplike varying substrate potential. We note
that here the scale of the $x$ axis is about three times larger than
in Fig.~\ref{f:step_complete}.}
\end{figure}

\begin{figure}
\begin{center}
\epsfig{file=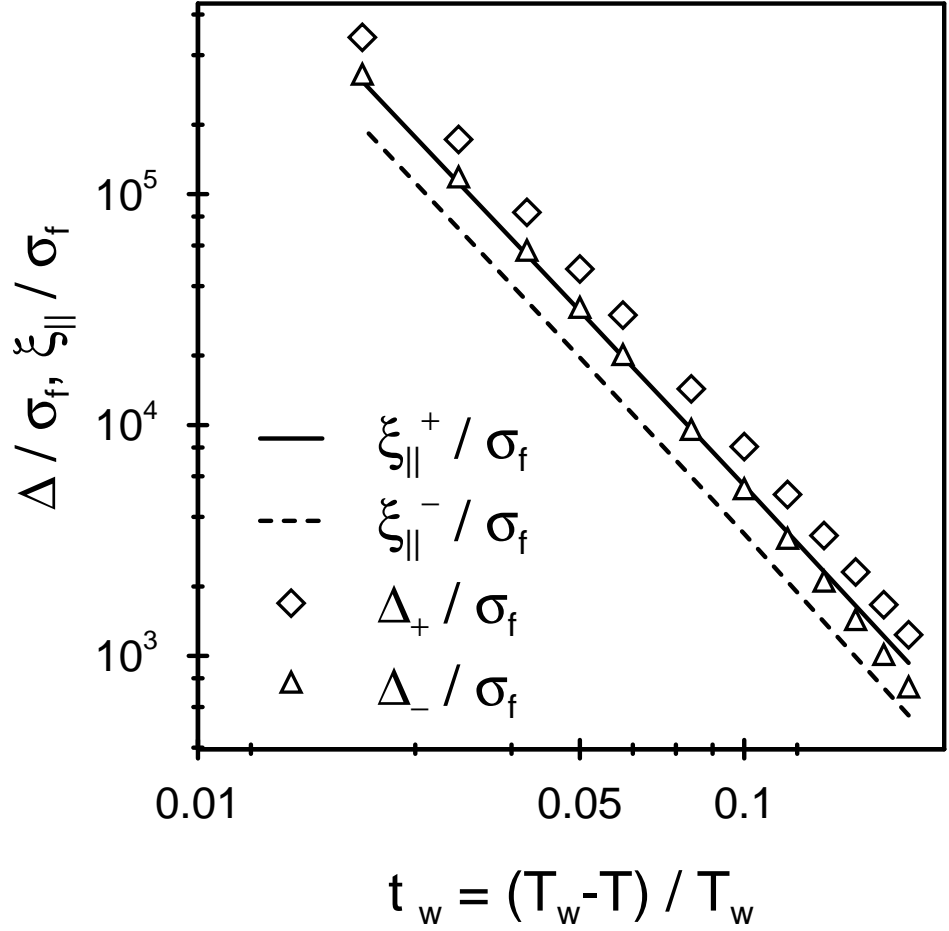, width=13cm, bbllx=70,
  bblly=315, bburx=515, bbury=770}
\end{center}
\caption{\label{f:step_intwidthcrit}
Widths $\Delta_{\pm}$ of the interface profiles ($\lozenge$, $\triangle$)
corresponding to the profiles shown in Fig.~\ref{f:step_critical} as
compared with the lateral height-height correlation lengths
$\xi_{\parallel}^{\pm}$ determined by $l_{\pm}$
(full and dashed line, Eq.~(\ref{e:step_corrlencrit})). $\Delta_{\pm}$
is defined as that value  
$|x|$ at which $l(x)$ starts to deviate from its respective asymptote $l_{\pm}$
by $10\%$ of $|l_+-l_-|$. Both $\Delta_{\pm}$ and
$\xi_{\parallel}^{\pm}$ diverge as $t_w^{-5/2}$ for $t_w\to0$ as
expected for critical wetting transitions.}
\end{figure}

\begin{figure}
\begin{center}
\epsfig{file=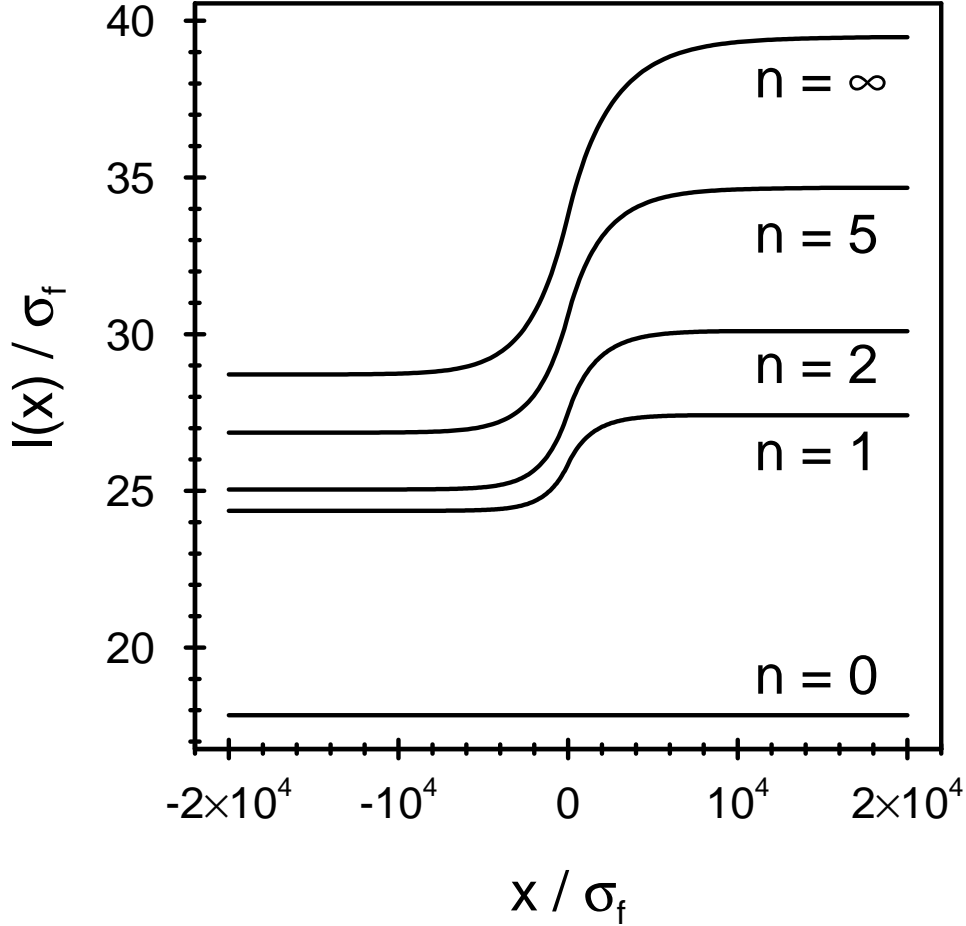, width=13cm, bbllx=80,
  bblly=320, bburx=525, bbury=770}
\end{center}
\caption{\label{f:2sl_diffd}
Liquid-vapor interface profiles on a LCS for different numbers
$n$ of monolayers in the inhomogeneous surface layer. $n=0$ corresponds to the
homogeneous substrate $w_H$, whereas $n=\infty$ leads to a SCS with
the substrate halves $w_+$ and $w_-$. $d_w$, $u_i^H$, and $u_{i,x}^H$ are chosen
equal to $d_w^-$, $u_i^-$, and $u_{i,x}^-$ as in Fig.~\ref{f:step_complete},
respectively, such that both substrate halves exhibit a common
critical wetting transition at $T_w^{+*}=T_w^{-*}=T_w^*=1.2$. The
interaction parameters of the surface layers are $u_i^+=1.1u_i^H$,
$u_i^-=1.05u_i^H$, and $g_x=g_z$. They do not affect the order of the
wetting transition and its transition temperature but they lead to a
difference between $l_+(T)$ and 
$l_-(T)$. The temperature is fixed at $T^*=1.1$, i.e., both substrate
halves are only partially wet at coexistence $\Delta\mu=0$. Moreover, we choose
$\Delta\mu^*=1.5\cdot10^{-6}$. As the number of monolayers in the
surface layer $n$ is increased the interface profile for a SCS
composed of $w_+$ and $w_-$ evolves out of the flat profile
for the homogeneous substrate $w_H$. Remarkably already a few
adsorbed monolayers have a pronounced effect on the wetting film.}
\end{figure}

\begin{figure}
\begin{center}
\epsfig{file=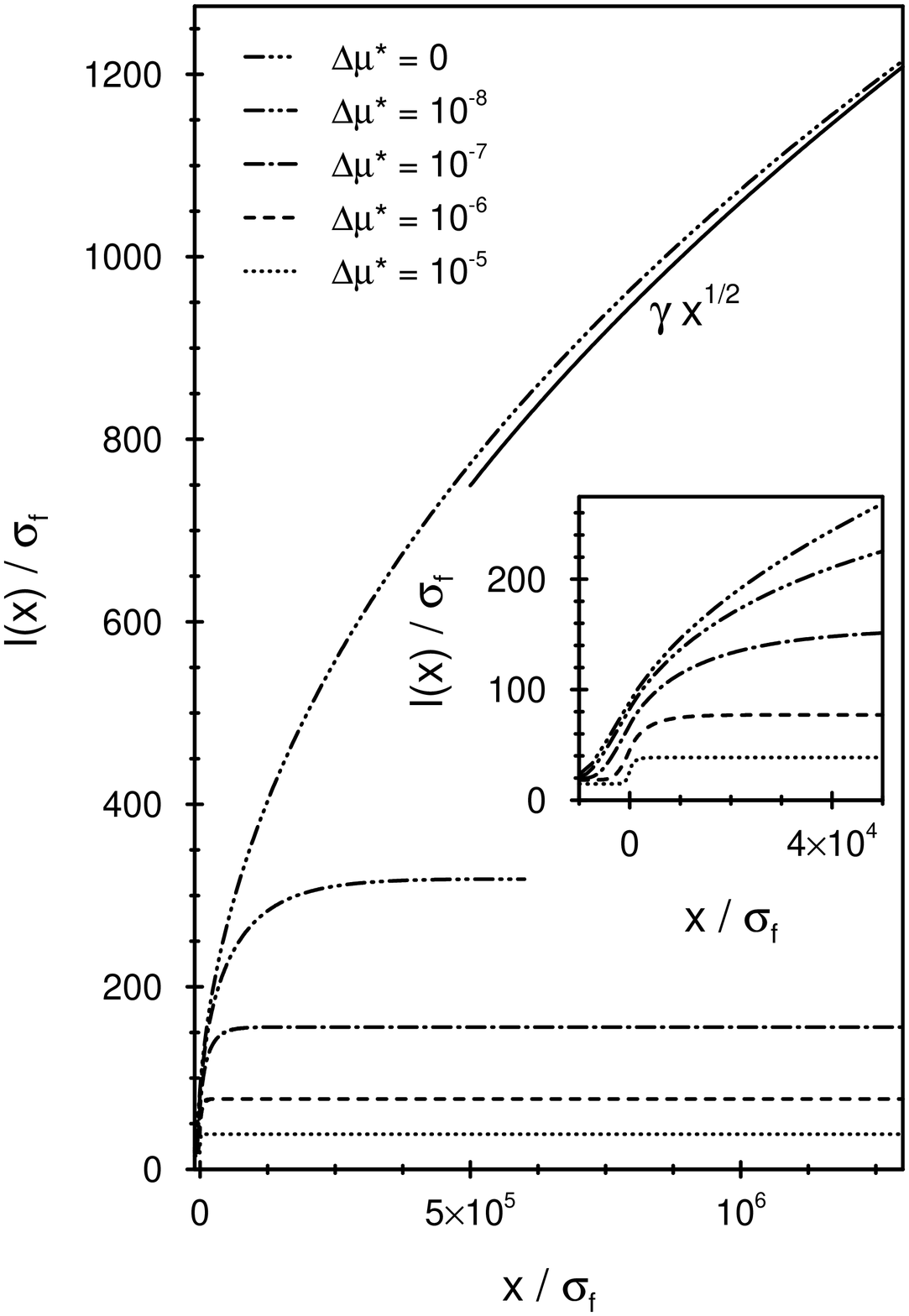, width=13cm, bbllx=30,
  bblly=35, bburx=545, bbury=785}
\end{center}
\caption{\label{f:2sl_partcomp}
Liquid-vapor interface profiles across a LCS with the substrate half $x>0$
completely wet and the substrate half $x<0$ only partially wet. The
parameters $d_w$, $u_i^H$, $u_i^+$, and $g_x/g_z$ are chosen as in
Fig.~\ref{f:2sl_diffd}, $u_i^-=0.75u_i^H$, and
we consider $n=10$ inhomogeneous monolayers. With this
choice of parameters the substrate half $x>0$ exhibits a critical wetting
transition at $T_w^{+*}=1.2$ whereas the substrate half $x<0$ undergoes a
weakly first-order 
wetting transition at $T_w^{-*} \approx 1.306$ with the thickness $l_-$ of the
liquidlike film of the order of $10\sigma_f$ in the vicinity of
$T_w^{-*}$. The temperature is fixed at $T^*=1.3$ such that at coexistence 
$\Delta\mu=0$ the substrate half $x<0$ remains partially wet. At coexistence $l(x)$
diverges according to $l(x\to\infty)=\gamma_{comp} x^{1/2}$ (see
Eqs.~(\ref{e:step_partcompcomp}) and
(\ref{e:step_partcompcompvdw})). As in the case of 
the SCS, the width $\Delta_+$ of the interface profile for $x>0$
diverges according to $(\Delta\mu)^{-2/3}$ for $\Delta\mu\to0$;
$\Delta_-$ remains finite 
even at $\Delta\mu=0$, for which it is of the order of
$\xi_{\parallel}^-\sim 10^4\sigma_f$. The inset provides a
magnified view of the region around $x=0$. For $\Delta\mu\to0$ and
$x\to\infty$ the profiles exhibit scaling and are governed by the same
scaling function $f(y)$ as the SCS (Eq.~(\ref{e:step_scalingcomp})).}
\end{figure}

\begin{figure}
\begin{center}
\epsfig{file=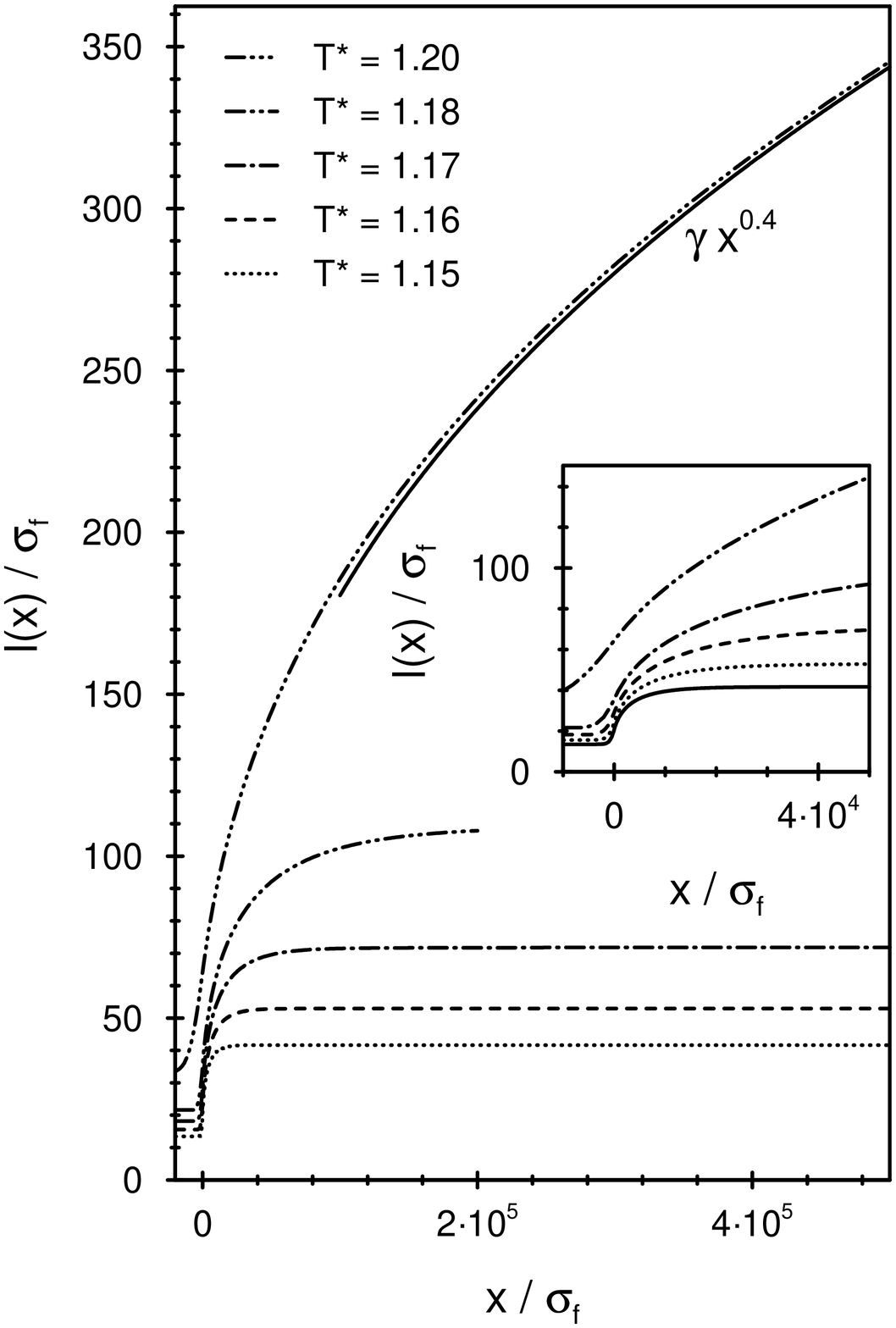, width=13cm, bbllx=30,
  bblly=35, bburx=545, bbury=785}
\end{center}
\caption{\label{f:2sl_partcrit}
Liquid-vapor interface profiles across a LCS along a thermodynamic
path at coexistence $\Delta\mu=0$ approaching the critical wetting
transition temperature $T_w^{+*}=1.2$ of the substrate half
$x>0$. Here we choose the same values of $d_w$, $u_i^H$, and $g_x/g_z$ as in
Figs.~\ref{f:2sl_diffd} and \ref{f:2sl_partcomp},
$u_i^+=u_i^H$, and $u_i^-=0.92u_i^H$. The latter
choice induces a weakly first-order wetting transition on the substrate half
$x<0$ at $T_w^*\approx1.216$ with a thickness of the liquidlike
wetting film of the order of $10\sigma_f$ slightly below the
transition temperature. In the limit $T\nearrow T_w^+$    
$l_+(T)$ diverges as $(T_w^+-T)^{-1}$ whereas $l_-(T)$
remains finite. At $T=T_w^+$ one has the power-law behavior $l(x\to\infty)=
\gamma_{crit}x^{2/5}$ (see Eqs.~(\ref{e:step_partcompcrit}) and
(\ref{e:step_partcompcritvdw})). As in
the case of the SCS, the width $\Delta_+$ of 
the interface for $x>0$ diverges as $t_w^{-5/2}$ with
$t_w=(T_w^+-T)/T_w^+\to0$, whereas 
$\Delta_-$ remains finite even at $t_w=0$, where it is of the order of
$\xi_{\parallel}^-\sim 10^4\sigma_f$. The inset provides a
magnified view of the region around $x=0$. For
$t_w\to0$ and $x\to\infty$ the profiles exhibit
scaling and are governed by the same scaling function $g(y)$ as the
SCS (Eq.~(\ref{e:step_scalingcrit})).}
\end{figure}

\begin{figure}
\begin{center}
\epsfig{file=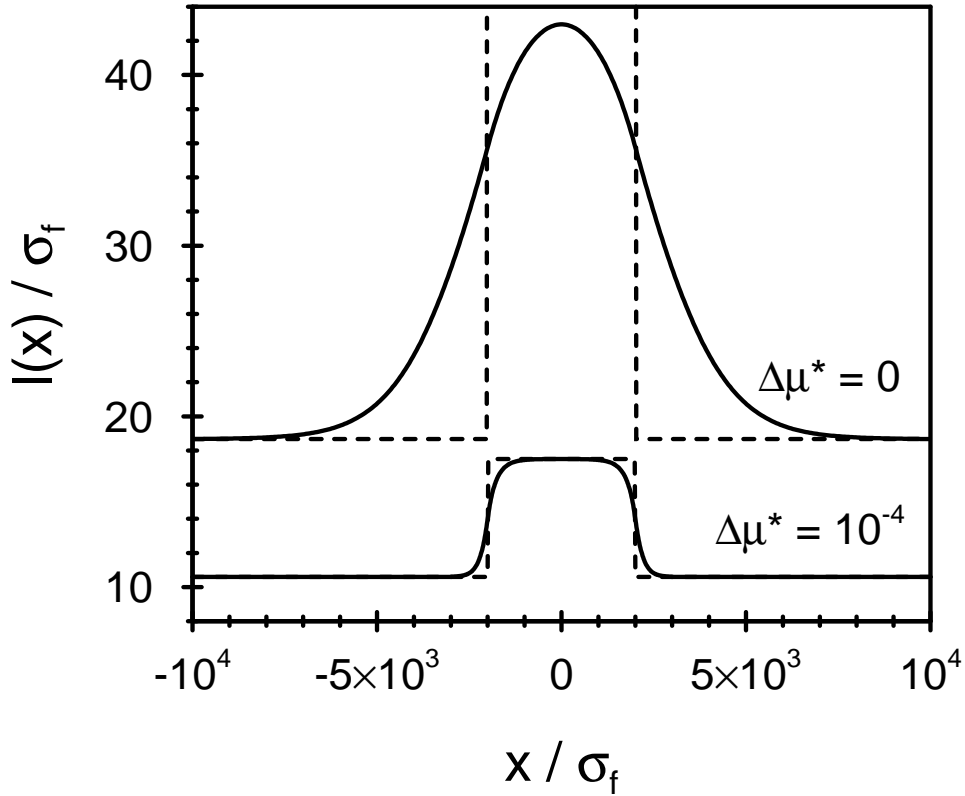, width=13cm, bbllx=70,
  bblly=315, bburx=520, bbury=720}
\end{center}
\caption{\label{f:stripe_complete}
Liquid-vapor interface profiles on a CST (full lines). The slab
$w_{st}$ is symmetric around $x=0$ and forms a chemical stripe at the
substrate surface with a width
$a/\sigma_f=4000$. The parameters for the substrate
potential are chosen as in Fig.~\ref{f:step_complete}, i.e., the whole
substrate undergoes a critical wetting transition at $T_w^*=1.2$. The
critical wetting transition at $T_w^*=1.0$ corresponding to the
slab $w_{st}$ is suppressed due to the 
finite lateral extension of the slab. The profiles correspond to the temperature
$T^*=1.1$ so that at coexistence $\Delta\mu=0$ the inhomogeneous
substrate as a whole is
only partially wet. Upon approaching coexistence $\mu\to\mu_0$ the
interface profile broadens. The dashed lines indicate the equilibrium
film thicknesses corresponding to the flat and homogeneous ``$+$'' and
``$-$'' substrate; $l_+=\infty$ for $\Delta\mu=0$. If the
width $\Delta_+$ of the transition regions near $|x|=a/2$ is small compared to
$a$ (as it is the case for $\Delta\mu^*=10^{-4}$) one has $l(x)\approx l_+$
in the middle of the stripe.} 
\end{figure}

\begin{figure}
\begin{center}
\epsfig{file=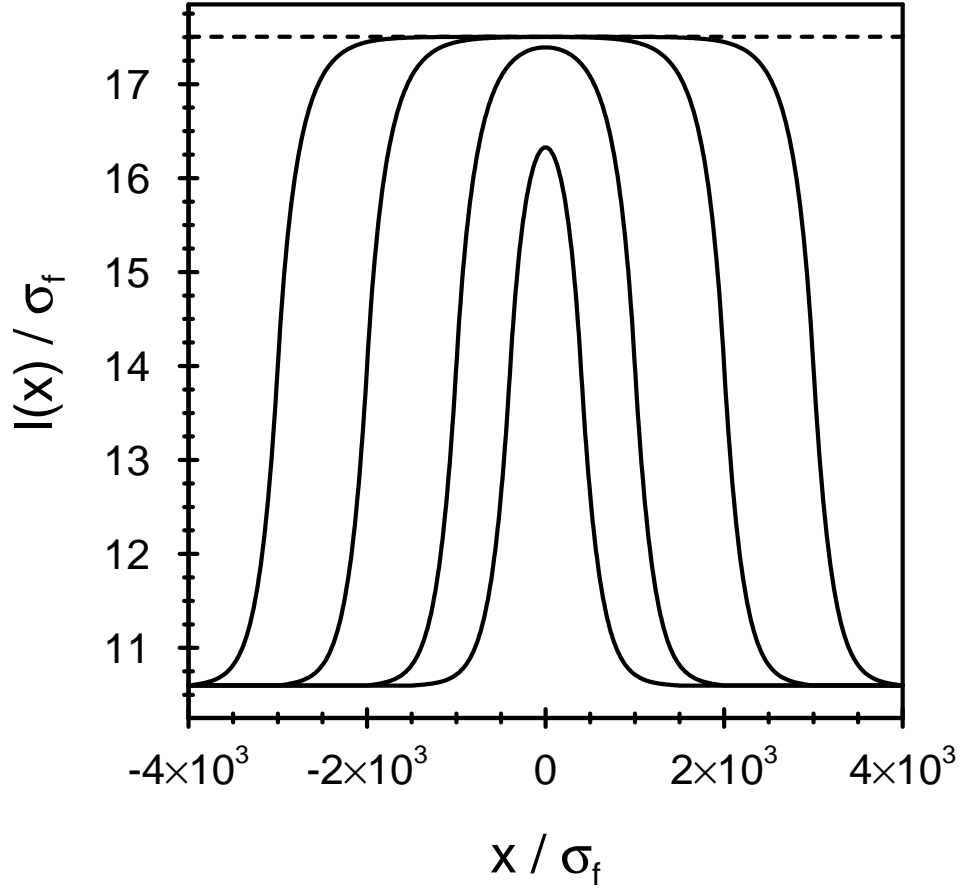, width=13cm, bbllx=70,
  bblly=325, bburx=535, bbury=775}
\end{center}
\caption{\label{f:stripe_diffwidth}
Change of the morphology of the liquid-vapor interface profiles on a
CST upon a variation of the stripe width $a$. The profiles correspond to
$a/\sigma_f=800$, $2000$, $4000$, and $6000$ from the inner to the
outer one. The temperature $T$ and the interaction potentials are
chosen as in Fig.~\ref{f:stripe_complete} and $\Delta\mu^*=10^{-4}$, i.e.,
the interface profile for $a=4000\sigma_f$ is the same as in
Fig.~\ref{f:stripe_complete}. If the stripe width is decreased the
region where $l(x)$ attains the equilibrium film thickness $l_+$
(indicated by the dashed line) shrinks and ultimately disappears. Only
for $a\gg a_0$, where $a_0\approx2\Delta_+(l_+)$ is a characteristic
crossover width, $l(x=0)$ attains the value $l_+$. One has
$\Delta_+\sim\xi_{\parallel}^+$ (compare
Fig.~\ref{f:step_intwidthcomp}) and $\xi_{\parallel}^+ =
\sqrt{\sigma_{lg}/6a_2^+}\,l_+^2$ in the limit $\Delta\mu\to0$, i.e.,
$l_+\to\infty$ (see Eq.~(\ref{e:step_corrlencomp})).
For the present system $\xi_{\parallel}^+\approx470\sigma_f$ and
$a_0\approx750\sigma_f$.}
\end{figure}

\begin{figure}
\begin{center}
\epsfig{file=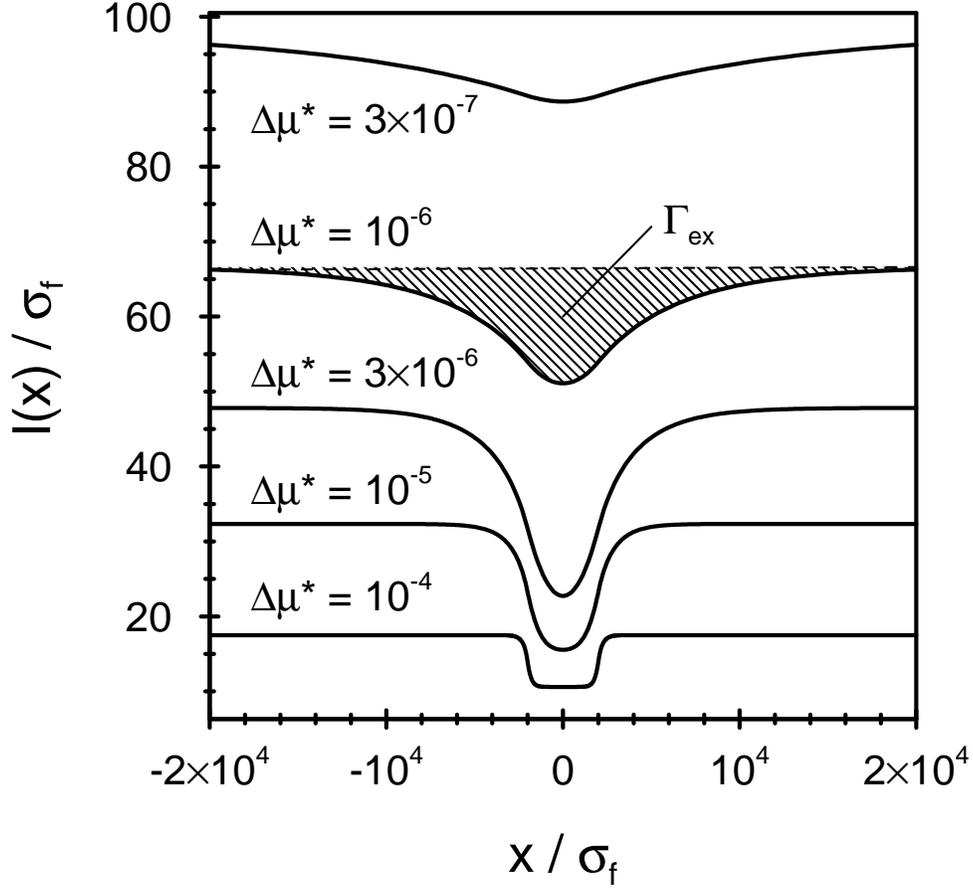, width=13cm, bbllx=65,
  bblly=325, bburx=535, bbury=775}
\end{center}
\caption{\label{f:stripe_exchanged}
Morphology of liquid-vapor interface profiles on a SST with the same
parameters as in Fig.~\ref{f:stripe_complete} but with the chemical
species of the stripe and the surrounding region exchanged such that
the slab $w_{st}$ favors \emph{thinner} liquidlike films than the
embedding substrate. Since the 
temperature $T^*=1.1$ is above the wetting transition temperature
corresponding to the outer region the substrate as a whole is completely wet at
coexistence $\Delta\mu=0$. Upon approaching coexistence the
equilibrium film thickness $l_-$ corresponding to the surrounding
substrate diverges and the ``dent'' induced by the presence of the
stripe is deepening but finally it is smeared out and vanishes. Upon
decreasing $\Delta\mu$ at first the depth $l_--l(x=0)$ of the ``dent'' 
increases, reaches a maximum value $(l_--l(0))_{max} \approx
25\sigma_f$ at $\Delta\mu^*\approx3\cdot10^{-6}$, and finally vanishes as
$(\Delta\mu)^{2/5}$ for $\Delta\mu\searrow0$. Since the width of the
dent is governed by the lateral correlation length
$\xi_{\parallel,comp}^-(\Delta\mu)$ which diverges as
$(\Delta\mu)^{-2/3}$ the depletion of the coverage
$|\Gamma_{ex}|\sim(l_--l(0))\,\xi_{\parallel,comp}^-$ induced by the stripe
\emph{diverges} as $(\Delta\mu)^{-4/15}$.}
\end{figure}

We have obtained the following main results:

\begin{enumerate}
\item Based on the description of wetting phenomena on homogeneous
  substrates within the framework of density functional theory
  (Sec.~\ref{s:theoryhomog}, Fig.~\ref{f:hom_sharpkink}), we have
  derived the systematic decomposition of the grand canonical
  potential of a fluid in contact with a chemically heterogeneous
  substrate into bulk, surface and line contributions. The minimum of
  the latter yields the equilibrium morphology of the adsorbed
  liquidlike wetting films and the associated line tension within mean
  field theory. As paradigmatic cases we have studied a simple
  chemical step (SCS, see Fig.~\ref{f:step_2sl_system} for
  $n\to\infty$ and Subsec.~\ref{s:step_theory}), a chemical step
  within a surface layer supported by a homogeneous substrate (LCS,
  see Fig.~\ref{f:step_2sl_system} and Subsec.~\ref{s:2sl_theory}),
  and a chemical stripe (CST, see Fig.~\ref{f:stripe_system} and
  Subsec.~\ref{s:stripe_theory}).
\item Across a SCS the profiles of the liquid-vapor interface of the
  liquidlike adsorbed wetting film morphology interpolate between
  their asymptotic values corresponding to the wetting film
  thicknesses on the two individual, homogeneous substrates forming
  the SCS (Figs.~\ref{f:step_complete} and \ref{f:step_critical}). The
  curvature of the profiles changes sign near the position of the step
  (Fig.~\ref{f:step_zeroline}). On each side of the step the lateral
  width $\Delta_{\pm}$ of this transition region is governed by the corresponding
  lateral correlation length $\xi_{\parallel}$ of the height-height
  correlation function (see Figs.~\ref{f:step_intwidthcomp} and
  \ref{f:step_intwidthcrit}, and Eqs.~(\ref{e:step_corrlencomp}) and
  (\ref{e:step_corrlencrit})). This width diverges according to power laws
  which depend on whether a complete or critical wetting transition
  (Fig.~\ref{f:step_partvscomp}) is approached. Near these transitions
  the dependence of the profiles on the lateral coordinate $x$ and on
  $\xi_{\parallel}$ exhibits scaling properties
  (Eqs.~(\ref{e:step_scalingcomp}) and (\ref{e:step_scalingcrit}))
  with singular scaling functions (Fig.~\ref{f:step_scalingfcts}) such
  that at the wetting 
  transitions the profiles diverge algebraicly as function of $x$ (see
  Eqs.~(\ref{e:step_partcompcomp}), (\ref{e:step_partcompcrit}), and
  Figs.~\ref{f:2sl_partcomp} and \ref{f:2sl_partcrit}). The
  corresponding amplitudes as well as those for the van der Waals
  tails of the profiles (see Eqs.~(\ref{e:step_vdwtcrit}) and
  (\ref{e:step_vdwtcomp})) have been determined analytically. The
  various power laws are modified by retardation
  (Subsec.~\ref{s:retardation}). Near first-order wetting transitions
  the interface profiles vary on a molecular scale
  (Fig.~\ref{f:step_1storder}) and are sensitive to the details of the
  laterally varying substrate potential.
\item The morphology of the wetting film across a chemical step within
  a surface layer (LCS) is similar to that on a SCS. The confinement
  of the chemical heterogeneity to a thin surface layer leads to more
  rapidly decaying van der Waals tails
  (Eqs.~(\ref{e:2sl_vdwt})--(\ref{e:2sl_vdwtcomp})) as compared with
  the SCS and modified amplitudes for the power-law divergences as
  function of $x$ (see the discussion after
  Eq.~(\ref{e:step_partcompcritvdw})). The scaling functions for the
  SCS and the LCS are the same. The main structural features of the wetting film
  across a LCS are already induced by very few heterogeneous surface
  monolayers on the substrate (Fig.~\ref{f:2sl_diffd}).
\item If the chemical heterogeneity is confined to a stripe (CST)
  complete wetting of the inner region is inhibited by an incompletely
  wetted outer region (Fig.~\ref{f:stripe_complete}). The leakage of
  the liquid into the outer region is governed by $\xi_{\parallel}$
  corresponding to the latter. Thus the tight confinement of the liquid to the
  stripe can be accomplished by choosing an embedding outer material
  on which only thin wetting films can form. This confinement can be
  achieved even if the inner material prefers macroscopicly thick
  wetting films. In the latter case and for wide stripes with width
  $a$ the interface profile has the shape of a semi-ellipse
  (Eq.~(\ref{e:stripe_ellipse})) and the excess coverage supported by
  the stripe scales as $a^{3/2}$
  (Fig.~\ref{f:stripe_confinement}). The perturbation of the interface
  profile near one step by the distant step has been determined
  analytically (Eq.~(\ref{e:stripe_dwallcorr})) as well as the
  corresponding van der Waals tails characterizing the decay into the
  outer region (Eqs.~(\ref{e:stripe_vdwtcrit}) and
  (\ref{e:stripe_vdwtcomp})). Only for sufficiently wide stripes the
  film thickness on the stripe can attain the
  value $l_+$ for the corresponding homogeneous case
  (Fig.~\ref{f:stripe_diffwidth}). If, on the other hand, the outer
  region undergoes a complete or critical wetting transition but the
  inner region does not favor it, the thickening wetting film spills over
  into the stripe region and drags its wetting behavior along leaving
  behind a dent (Fig.~\ref{f:stripe_exchanged}). The depth of the dent
  behaves nonmonotonously as function of the undersaturation
  $\Delta\mu$. Ultimately, for $\Delta\mu\to0$ the depth of the dent
  vanishes and its width diverges proportional to $\xi_{\parallel}$ so
  that the net depletion of the coverage caused by the stripe diverges
  as $(\Delta\mu)^{-4/15}$ (Fig.~\ref{f:stripe_exchanged}).
\item The dependence of the line tension $\tau_{CST}(a)$ associated
  with a chemical stripe on the stripe width $a$ leads to a lateral force per
  unit length $f(a) = -\partial\tau_{CST}(a)/\partial a$
  (Eqs.~(\ref{e:stripe_force1}) and (\ref{e:stripe_force2})) acting on the stripe
  and leading to a compression or dilation which is balanced by the
  elastic forces of the substrate. Except for small stripe widths
  $a\sim\sigma_f$ the constant contribution $f_0=\sigma_{w_-g}-\sigma_{w_+g}$
  is the dominating one. The excess contribution $f_{ex}(a) =
  f(a)-f_0(a)$ decays as $f_{ex}(a\to\infty)\sim a^{-3}$
  (Fig.~\ref{f:stripe_force}). Whereas for solid substrates this force
  has practically no effect we expect that for liquidlike wetting
  films on a liquid substrate decorated with a Langmuir-Blodgett film
  which contains a stripe of different material the force can become
  important. In this case the adsorbed wetting films cause 
  a detectable change in the stripe width depending on the compressibility of the
  Langmuir-Blodgett films. 
\item In Sec.~\ref{s:general} we have presented a systematic
  microscopic derivation of the Euler-Lagrange equation
  (Eq.~(\ref{e:generalele})) for a liquidlike film adsorbed on an
  arbitrary chemically structured substrate
  (Fig.~\ref{f:generalinh}). This approach also provides a microscopic
  calculus for determining local, laterally varying wall-vapor and
  wall-liquid surface tensions (Eqs.~(\ref{e:generaldepwlstens}) and
  (\ref{e:generalwallgastens})) required for macroscopic
  analyses. Lateral inhomogeneities within the repulsive part of the
  substrate potential modify these expressions with respect to what is
  expected intuitively (see Eqs.~(\ref{e:localeffintpot}) and
  (\ref{e:generalwallgastens})).
\end{enumerate}

\acknowledgements

We gratefully acknowledge financial support by the German Science
Foundation within the special research initiative
\emph{Wetting and Structure Formation at Interfaces}.

\appendix

\section{Potentials of heterogeneous substrates}
\label{s:app_potentials}

Approximately we determine the substrate potential by a pairwise
summation of the interaction between a single fluid particle and all
substrate particles. To this end we assume that the molecules in the
substrate are located at orthorhombic lattice sites with the lattice
constants $g_x$, $g_y$, and $g_z$ in the $x$, $y$, and $z$ direction,
respectively, which are taken to be constant throughout the substrate. The leading
contribution from this summation corresponds to a  
three-dimensional integration over the substrate volume. The discrete
sum generates in addition subdominant contributions which are proportional
to powers of the lattice spacings. The substrate potential can be
written as a sum $V(x,z)=V_{att}(x,z)+V_{rep}(x,z)$ of an attractive
and a repulsive contribution. In the attractive contribution we take
into account the two leading orders of this power series
whereas the repulsive contribution is modeled by a steplike crossover
between the repulsive parts of the potentials of the corresponding
homogeneous, semi-infinite substrates. This assumption is justified
because the repulsive interaction decays very rapidly and is only
significant for $z \lesssim 1.5\sigma_f$ where $z=0$ denotes the
position of the nuclei of the top substrate layer.

The different chemical species are distinguished by ``$+$'' and ``$-$''
denoting the different potential coefficients. If a homogeneous
substrate is covered by a heterogeneous surface layer of finite
thickness we denote its constituent molecules by ``$H$''. For the
interparticle potential we adopt the Lennard-Jones form
\begin{equation}
\phi_{\pm,H}(r) =
4\epsilon_{\pm,H}\left[\left(\frac{\sigma_{\pm,H}}{r}\right)^{12} -
  \left(\frac{\sigma_{\pm,H}}{r}\right)^6\right].
\end{equation}

Under these conditions one finds for a simple chemical step (SCS),
i.e., for two adjacent quarter spaces $w_+$ and $w_-$ (see
Fig.~\ref{f:step_2sl_system} with $n=\infty$)
\begin{eqnarray}\label{e:app_step_potatt}
V_{att}^{SCS}(x,z) & = & -\frac{u_3^++u_3^-}{2}\frac{1}{z^3} +
\frac{u_3^+-u_3^-}{2}
\left(\frac{1}{x^3}-\left(\frac{r}{xz}\right)^3+\frac{3}{2}\frac{1}{xzr}\right)
\nonumber\\
& & -\frac{u_4^++u_4^-}{2}\frac{1}{z^4} -
\frac{u_4^+-u_4^-}{2}
\left(\frac{x}{z^4r}+\frac{1}{2}\frac{x}{z^2r^3}\right) \nonumber\\
& & + \frac{u_{4,x}^+-u_{4,x}^-}{2}
\left(\frac{1}{x^4}-\left(\frac{z}{x^4r}+\frac{1}{2}\frac{z}{x^2r^3}\right)\right)
+ {\mathcal O}(x^{-m}z^{-n},\,m+n=5)
\end{eqnarray}
with $r = \sqrt{x^2+z^2}$ and
\begin{equation}\label{e:app_step_potrep}
V_{rep}^{SCS}(x,z) = \Theta(-x)\frac{u_9^-}{z^9} +
\Theta(x)\frac{u_9^+}{z^9}.
\end{equation}
The coefficients $u_3^{\pm}$, $u_4^{\pm}$, 
and $u_9^{\pm}$ are defined by the potentials of the respective
homogeneous, flat, semi-infinite substrates $w_+$ and $w_-$
(see Eq.~(\ref{e:hom_substpot})), whereas $u_{4,x}^{\pm} =
u_4^{\pm}g_x/g_z$.

Similarly, for the homogeneous, flat substrate $w_H$ covered by two
different, adjacent surface layers $w_{S,\pm}$ which consist of $n=d+1$
monolayers (compare Fig.~\ref{f:step_2sl_system}) we obtain
\begin{eqnarray}\label{e:app_2sl_potatt}
V_{att}^{LCS}(x,z) & = & -\frac{u_3^H}{(z+(d+1)g_z)^3} -
\frac{u_4^H}{(z+(d+1)g_z)^4} \nonumber\\
& & - \frac{u_3^++u_3^-}{2}\left(\frac{1}{z^3}-\frac{1}{(z+dg_z)^3}\right) -
\frac{u_4^++u_4^-}{2}\left(\frac{1}{z^4}+\frac{1}{(z+dg_z)^4}\right)
\nonumber\\
& & + \frac{u_3^+-u_3^-}{2}
\left[\frac{2x^4+x^2\,\tilde{z}^2+2\tilde{z}^4}
  {2x^3\,\tilde{z}^3\,\tilde{r}}-
\frac{2x^4+x^2\,z^2+2z^4}{2x^3\,z^3\,r}\right] \nonumber\\
& & - \frac{u_4^+-u_4^-}{2}
\left[\frac{x\,(3\tilde{z}^2+2x^2)}{2\tilde{z}^4\,\tilde{r}^3}
  + \frac{x\,(3z^2+2x^2)}{2z^4\,r^3}\right] \nonumber\\
& & + \frac{u_{4,x}^+-u_{4,x}^-}{2}
\left[\frac{\tilde{z}\,(3x^2+2\tilde{z}^2)}{2x^4\,\tilde{r}^3}
  - \frac{z\,(3x^2+2z^2)}{2x^4\,r^3}\right] + {\mathcal
  O}(x^{-m}z^{-n},\,m+n=5)
\end{eqnarray}
where $\tilde{z} = (z+dg_z)$, $r = \sqrt{x^2+z^2}$, $\tilde{r} =
\sqrt{x^2+\tilde{z}^2}$, and
\begin{equation}\label{e:app_2sl_potrep}
V_{rep}^{LCS}(x,z) \equiv V_{rep}^{LCS}(z) = \frac{u_9^H}{z^9}.
\end{equation}
The coefficients $u_j^{\pm}$ and $u_j^H$ correspond to the
coefficients of the homogeneous, flat substrate covered by a homogeneous
surface layer (see Eqs.~(\ref{e:sl_potatt}) and (\ref{e:sl_potrep}))
whereas $u_{4,x}^{H,\pm}=u_4^{H,\pm}g_x/g_z$.

Finally, for the slab $w_{st}$ immersed in a homogeneous
substrate $w$ (see Fig.~\ref{f:stripe_system}) the summation yields
\begin{eqnarray}\label{e:app_stripe_potatt}
V_{att}^{CST}(x,z) & = & -\frac{u_3^-}{z^3}-\frac{u_4^-}{z^4} -
\frac{u_3^+-u_3^-}{2}\left(\frac{1}{X^3}-\frac{1}{Y^3}\right)
- \frac{u_{4,x}^+-u_{4,x}^-}{2}
\left(\frac{1}{X^4}+\frac{1}{Y^4}\right) \nonumber\\
& & + \frac{u_3^+-u_3^-}{4}\left
  [ \frac{2X^4+X^2\,z^2+2z^4}{z^3\,X^3\,(X^2+z^2)^{1/2}} -
  \frac{2Y^4+Y^2\,z^2+2z^4}{z^3\,Y^3\,(Y^2+z^2)^{1/2}} \right]
\nonumber\\
& & + \frac{u_4^+-u_4^-}{4}\left
  [ \frac{X\,(3z^2+2X^2)}{z^4\,(X^2+z^2)^{3/2}} -
  \frac{Y\,(3z^2+2Y^2)}{z^4\,(Y^2+z^2)^{3/2}} \right] \nonumber\\
& & + \frac{u_{4,x}^+-u_{4,x}^-}{4}\left
  [ \frac{z\,(3X^2+2z^2)}{X^4\,(X^2+z^2)^{3/2}} +
  \frac{z\,(3Y^2+2z^2)}{Y^4\,(Y^2+z^2)^{3/2}} \right] \nonumber\\
& & + {\mathcal O}(x^{-m}z^{-n},\,m+n=5)
\end{eqnarray}
where the abbreviations $X$ and $Y$ denote $X=x-a/2$ and
$Y=x+a/2$, and
\begin{equation}\label{e:app_stripe_potrep}
V_{rep}^{CST}(x,z) = \Theta\left(|x|-\frac{a}{2}\right)\,\frac{u_9^-}{z^9} +
\Theta\left(\frac{a}{2}-|x|\right)\,\frac{u_9^+}{z^9}
\end{equation}
with the coefficients defined as above for the SCS.

At large separations between a fluid particle and the substrate
retardation effects set in so that $\phi_{\pm,att}(r\to\infty) \sim
r^{-7}$ \cite{israelachvili}. In this case the leading contribution to
$V_{att}^{SCS}(x,z)$ is 
\begin{eqnarray}\label{e:app_step_potattret}
V_{att}(x,z) & = & -\frac{v_4^++v_4^-}{2\,z^4} + 
\frac{v_4^+-v_4^-}{2}\left\{\frac{1}{x^4} -
  \frac{2}{\pi}\left(\frac{\arctan(x/z)}{z^4} 
    + \frac{\arctan(z/x)}{x^4}\right) \right. \nonumber\\
& & \left. + \frac{2}{3\pi}\,\frac{x^2-3z^2}{z^3\,x^3} -
\frac{8}{3\pi}\,\frac{x}{z^3\,(x^2+z^2)} \right\} + {\mathcal
O}(x^{-m}z^{-n},\,m+n=5).
\end{eqnarray}

\section{Decomposition of the grand canonical density functional}
\label{s:app_subdiv}

\subsection{Single chemical step}

The sharp-kink ansatz (Eq.~(\ref{e:step_sharpkink})) for
Eq.~(\ref{e:ldafunctional}) leads to the decomposition of the grand
canonical free energy given by Eqs.~(\ref{e:step_subdiv}),
(\ref{e:bfedensity}), and (\ref{e:step_sfe}) with
\begin{equation}\label{e:app_step_sfe}
\Omega_s^{\pm}(l_{\pm}) = l_{\pm}\Delta\Omega_b + \sigma_{w_{\pm}l} +
\sigma_{lg} +  \omega(l_{\pm}),
\end{equation}
where the effective interface potentials $\omega_{\pm}(l)$ are given
by (compare with Eq.~(\ref{e:hom_effintpot}))
\begin{equation}\label{e:app_step_effintpot}
\omega_{\pm}(l) = \Delta\rho\left\{\rho_l\int\limits_{l-d_w^{\pm}}^{\infty}
  dz\,t(z) - \int\limits_l^{\infty}dz V_{\pm}(z) \right\}.
\end{equation}
Here we have omitted artificial contributions generated by truncating
the system; these contributions are discussed in Ref.~\cite{koch}. The
surface contribution is that of two half
substrates $w_+$ and $w_-$ covered by liquidlike wetting layers of
thickness $l_+$ and $l_-$ which are exposed to the potentials
$V_+(z)$ and $V_-(z)$, respectively, of the corresponding semi-infinite, homogeneous
substrates. This defines a reference system such that the deviation
of the smooth profile $l(x)$ from the reference configuration (see the
dashed lines in Figs.~\ref{f:step_2sl_system} and \ref{f:stripe_system})
due to the heterogeneity and due to the smooth crossover of $V(x,z)$ from
$V_-(z)$ to $V_+(z)$ leads to the line contribution
\begin{equation}
\Omega_l[l(x)] = \tau(d_w^{\pm},l_{\pm}) + \tilde{\omega}[l(x)].
\end{equation}
The term $\tau(d_w^{\pm},l_{\pm})$, which is independent of $l(x)$, is given
in Eqs.~(B12), (B19), and (B20) in Ref.~\cite{koch}. (We note that there
the $+$ sign of the last term in Eq.~(B20) must be replaced by a $-$
sign.) The contribution which depends functionally on $l(x)$ reads  
\begin{eqnarray}\label{e:app_step_taudep}
\tilde{\omega}[l(x)] & = &
\Delta\Omega_b\Gamma_{ex} + \int\limits_{-\infty}^{\infty} dx
\{\omega(x,l(x);d_w^+)-\omega(x,l_{\infty}(x);d_w^+)\} \nonumber\\
& & - \Delta\rho\,\rho_l \left( \int\limits_{-\infty}^{\infty} dx
\int\limits_{l(x)-d_w^+}^{l_{\infty}(x)-d_w^+}dz \, \bar{t}(x,z) - 
\int\limits_{-\infty}^{\infty} dx
\int\limits_{l(x)-d_w^-}^{l_{\infty}(x)-d_w^-}dz \, \bar{t}(x,z) \right)
\nonumber\\
& & - \frac{1}{2} (\Delta\rho)^2 \int\limits_{-\infty}^{\infty} dx
\int\limits_{-\infty}^{\infty} dx' \int\limits_0^{\infty} dz
\int\limits_0^{l(x)-l(x')} dz' \, \bar{w}(x-x',z-z')
\end{eqnarray}
where $l_{\infty}(x)=l_-\Theta(-x)+l_+\Theta(x)$.
The first term measures the cost in free energy for
replacing a certain volume of vapor by the liquid phase. It
is proportional to the excess coverage
\begin{equation}
\Gamma_{ex} = \int\limits_{-\infty}^{\infty} dx (l(x)-l_{\infty}(x))
\end{equation}
and vanishes at coexistence $\mu=\mu_0$. The second term
is the integrated ``local'' effective interface potential 
\begin{equation}\label{e:app_localeffintpot}
\omega(x,l;d_w) = 
\Delta\rho\left\{\rho_l\int\limits_{l-d_w}^{\infty} dz \, t(z)
-\int\limits_l^{\infty} dz \, V(x,z) \right\}.
\end{equation}
The third term involving integrals of
\begin{equation}\label{e:app_txz}
\bar{t}(x,z) = \int\limits_x^{\infty}dx'\int\limits_z^{\infty}dz'\,
\bar{w}(x',z')
\end{equation} 
takes into account the difference between $d_w^-$ and $d_w^+$ and
vanishes for $d_w^- = d_w^+$. The last contribution with
\begin{equation}\label{e:app_wbar}
\bar{w}(x,z) =
\int\limits_{-\infty}^{\infty}dy\,\tilde{w}(\sqrt{x^2+y^2+z^2})
\end{equation}
describes the free energy of the 
deformed free liquid-vapor interface. It is a \emph{nonlocal} functional of
$l(x)$. A gradient expansion to the first order of this
contribution yields the fourth term of the corresponding \emph{local}
functional $\tilde{\omega}_{loc}[l(x)]$:
\begin{eqnarray}\label{e:app_locfunc}
- \frac{1}{2} (\Delta\rho)^2 & & \int\limits_{-\infty}^{\infty} dx
\int\limits_{-\infty}^{\infty} dx' \int\limits_0^{\infty} dz
\int\limits_0^{l(x)-l(x')} dz' \, \bar{w}(|x-x'|,|z-z'|) \nonumber\\
& & \longrightarrow \quad \sigma_{lg}\int\limits_{-\infty}^{\infty}dx\,
\left\{\sqrt{1+\left(\frac{dl}{dx}\right)^2} - 1\right\}.
\end{eqnarray}

\subsection{Chemical stripe}

The sharp-kink approximation of the density profile for the CST is
\begin{eqnarray}
\hat{\rho}(x,z) & = &
\left\{\Theta\left(|x|-\frac{a}{2}\right)\,\Theta(z-d_w^-) + 
\Theta\left(\frac{a}{2}-|x|\right)\,\Theta(z-d_w^+) \right\} \nonumber\\
& & \times \left[\Theta(l(x)-z)\rho_l + \Theta(z-l(x))\rho_g \right].
\end{eqnarray}
Insertion of $\hat{\rho}(x,z)$ into Eq.~(\ref{e:ldafunctional}) leads to
the decomposition
\begin{eqnarray}
\Omega([\hat\rho(x,z)];T,\mu;[\tilde{w}],[V]) & = &
\Lambda\,\Omega_b(\rho_g,T,\mu)
+ A\,\Omega_s(l_-;T,\mu;[\tilde{w}],[V])
\nonumber\\
& & + L_y\,\Omega_l([l(x)];T,\mu;[\tilde{w}],[V]).
\end{eqnarray}
with $\Omega_b$ given by Eq.~(\ref{e:bfedensity}). The surface
contribution stems from the reference system which in the case of the CST
is the structured substrate covered by a liquidlike layer of
thickness $l_-$ that is exposed to the potential $V_-(z)$:
\begin{equation}
\Omega_s(l_-) = \Omega_s^-(l_-)
\end{equation}
with $\Omega_s^-(l_-)$ defined by Eqs.~(\ref{e:app_step_sfe}) and
(\ref{e:app_step_effintpot}). 

The line contribution for the CST is given by
\begin{equation}\label{e:app_stripe_lcont}
\Omega_l[l(x)] = \tau(d_w^{\pm},l_-) + \tilde{\omega}[l(x)]
\end{equation}
and represents the free energy associated with the deviation of $l(x)$
from the asymptotic value $l_-$, i.e., of the reference configuration.
In Eq.~(\ref{e:app_stripe_lcont}) the first term does not depend on $l(x)$:
\begin{eqnarray}\label{e:app_stripe_tauindep}
\tau(d_w^{\pm},l_-) & = & -a(d_w^+-d_w^-)\Omega_b^{(l)} -
(d_w^+-d_w^-)\rho_l^2\int\limits_0^{\infty}dz\,t(z)
- a\rho_l\int\limits_{d_w^-}^{d_w^+}dz\,V_{-}(z) \nonumber\\
& & +2\rho_l^2 \left\{
\int\limits_0^{\infty}dx\,\int\limits_0^{\infty}dz
- \int\limits_0^{\infty}dx\,\int\limits_{d_w^+-d_w^-}^{\infty}dz
- \int\limits_{a}^{\infty}dx\,\int\limits_0^{\infty}dz
+ \int\limits_{a}^{\infty}dx\,\int\limits_{d_w^+-d_w^-}^{\infty}dz
\right\} \bar{t}(x,z) \nonumber\\
& & +\rho_l\int\limits_{-\infty}^{\infty}dx\,
\int\limits_{d_w^{\infty}(x)}^{\infty}dz\, \delta V(x,z)
-\Delta\rho\int\limits_{-\infty}^{\infty}dx\,
\int\limits_{l_-}^{\infty}dz\,\delta V(x,z)
\end{eqnarray}
with $\delta V(x,z) = V(x,z) - V_-(z)$ and
$d_w^{\infty}(x) = \Theta(|x|-a/2)d_w^-+\Theta(a/2-|x|)d_w^+$. The
interpretation of the different terms in 
Eq.~(\ref{e:app_stripe_tauindep}) is analogous to the SCS (compare
Ref.~\cite{koch}). The second term in Eq.~(\ref{e:app_stripe_lcont})
depends on $l(x)$:
\begin{eqnarray}\label{e:app_stripe_taudep}
\tilde{\omega}[l(x)] & = &
\Delta\Omega_b\Gamma_{ex}^{CST}
+ \int\limits_{-\infty}^{\infty} dx
\{\omega(x,l(x);d_w^-)-\omega(x,l_-;d_w^-)\} \nonumber\\
& & + \Delta\rho\,\rho_l \int\limits_{-\infty}^{\infty}dx
\int\limits_{-a/2}^{a/2}dx'
\int\limits_{l(x)-d_w^+}^{l(x)-d_w^-}dz \int\limits_{z}^{\infty}dv\,
\bar{w}(x-x',v) \nonumber\\
\nonumber\\
& & - \frac{1}{2} (\Delta\rho)^2 \int\limits_{-\infty}^{\infty} dx
\int\limits_{-\infty}^{\infty} dx' \int\limits_0^{\infty} dz
\int\limits_0^{l(x)-l(x')} dz' \, \bar{w}(x-x',z-z')
\end{eqnarray}
with
\begin{equation}
\Gamma_{ex}^{CST} = \int\limits_{-\infty}^{\infty} dx (l(x)-l_-).
\end{equation}
The interpretation of the first two terms is
the same as for the SCS; the third term is generated by the difference
of the excluded volumes at the surfaces of the inner and the outer
region and vanishes if $d_w^+=d_w^-$. The last term is the same as in
Eq.~(\ref{e:app_step_taudep}). 

The corresponding local functional expression
$\tilde{\omega}_{loc}[l(x)]$ follows by replacing the last term in
Eq.~(\ref{e:app_stripe_taudep}) by the local functional given in
Eq.~(\ref{e:app_locfunc}). We note that for all three substrate types
(SCS, LCS, and CST) the functionals $\tilde{\omega}[l(x)]$ (and therefore
also $\tilde{\omega}_{loc}[l(x)]$) exhibit the same structure if
$d_w^+ = d_w^-$.

\pagebreak

\end{document}